\documentclass[showpacs,preprintnumbers,amsmath,amssymb,nofootinbib]{revtex4}

\usepackage{color}   

\usepackage{graphicx}
\usepackage{dcolumn}
\usepackage{color}

\usepackage{graphicx}
\usepackage{dcolumn}
\usepackage{bm}

\usepackage[utf8]{inputenc}
\usepackage[T1]{fontenc}
\usepackage{ulem}
\usepackage{amsmath}
\usepackage{amssymb}
\usepackage{float}

\begin{document}

\def\bb    #1{\hbox{\boldmath${#1}$}}
\def\bb    #1{\hbox{\boldmath${#1}$}}

\def\blambda{{\hbox{\boldmath $\lambda$}}}
\def\eeta{{\hbox{\boldmath $\eta$}}}
\def\bxi{{\hbox{\boldmath $\xi$}}}
\def\bzeta{{\hbox{\boldmath $\zeta$}}}
\def\sD{D \!\!\!\!/}
\def\sd{\partial \!\!\!\!/}
\def\EQ{{\hbox{\boldmath $Eh.(\ref$}}}

\def\qcdu{{{}_{ \rm QCD}}}   
\def\qedu{{{}_{\rm QED}}}   
\def\qcdd{{{}^{ \rm QCD}}}   
\def\qedd{{{}^{\rm QED}}}   
\def\qcd{{{\rm QCD}}}   
\def\qed{{{\rm QED}}}   
\def\2d{{{}_{\rm 2D}}}         
\def\4d{{{}_{\rm 4D}}}         
\def\sg#1{ {\rm \,sign}(#1)\, }

\large

\title{ 
Possible Evidence for Neutral 
Color-Singlet  $q\bar q$ Quark Matter from High-Energy Pb-Emulsion Collisions }

\author{Cheuk-Yin Wong}

\affiliation{Physics Division, Oak Ridge National Laboratory\footnote{
    This research  has been supported in part by UT-Battelle, LLC,
    under contract DE-AC05-00OR22725 with the US Department of Energy
    (DOE). The US government retains and the publisher, by accepting
    the article for publication, acknowledges that the US government
    retains a nonexclusive, paid-up, irrevocable, worldwide license to
    publish or reproduce the published form of this manuscript, or
    allow others to do so, for US government purposes. DOE will
    provide  public access to these results of federally sponsored
    research in accordance with the DOE Public Access Plan
    (http://energy.gov/downloads/doe-public-access-plan), Oak Ridge,
    Tennessee, 37831 USA }\!, Oak Ridge, TN
37831, U.S.A.}

\begin{abstract}

The invariant mass spectrum of $e^+e^-$ pairs produced in high-energy Pb-emulsion collisions at 160 A GeV at
CERN SPS exhibits a highly complex structure of  a broad enhancement at 11$\pm$1  MeV with a full width of about 10 MeV, and  additional 
many narrow resonances within the experimental bin width of 2 MeV, including
a    prominent narrow  resonance at 19 $\pm$1 MeV  that provides an independent support
for the hypothetical X17 particle. We show that the highly complex spectrum may be coherently described as the signatures for the neutral color-singlet $q\bar q$ quark matter in both its deconfined phase as well as its
confined phase.  That is,  the broad enhancement at 11$\pm$1  MeV may arise from thermal annihilation of QED(U(1))-deconfined quarks and antiquarks into $e^+e^-$ pairs at 
the phase transition  temperature $T_c$(QED), which can be theoretically 
estimated to be 4.75 $\pm$ 1.04 MeV from the transitional equilibrium condition.
The observed   narrow  resonances at 3$\pm$1 and 7$\pm$1 MeV
 may correspond to the  QED(U(1))-deconfined    $u\bar u$  and  $d\bar d$ Coulomb bound states  near their quark rest masses, respectively,  whereas the observed  
narrow resonance at  19 $\pm$ 1 MeV  may correspond to the QED(U(1))-confined isoscalar QED meson.   The approximate agreement  
between the theoretical  and the experimental spectrum suggests tentatively that both QED(U(1))-confined and QED(U(1))-deconfined
neutral color-singlet $q\bar q$ quark matter may have been produced in these high-energy Pb-emulsion collision, pending confirmation of the Pb-emulsion collision data.   We propose future
experiments to confirm or refute these findings.

\end{abstract}


\maketitle

\section{Introduction}
In search of new neutral boson particles with masses below 100 MeV
decaying into $e^+e^-$ pairs, Jain and Singh examined the tracks of
$e^+e^-$ pairs produced in central collisions of 
high-energy $^{207}$Pb nuclei  incident on the AgBr emulsion at the energy of 160
A GeV at CERN SPS \cite{Jai07}.  Among their data sets, they studied
in particular the invariant mass spectrum of the $e^+e^-$ pairs   
originating at a distance of 50 $\mu$m to 200 $\mu$m away from the
primary Pb-emulsion collision vertex, corresponding to the search for
neutral boson particles which were produced in the Pb-emulsion
collisions and subsequently decayed into $e^+e^-$ pairs with lifetimes
of order $10^{-15}$s to $10^{-12}$s.  Remarkably, they observed in
2007 a highly complex structure of many narrow resonances some of which  rest on top of a broad  enhancement \cite{Jai07}
as shown in Fig.\ 1, with  the  experimental resonance energies  tabulated in Column 6 of   
 Table 1.  The  observed broad enhancement and  the narrow resonances are anomalous
because their invariant masses, lying below the pion mass but above the
photon mass, place them outside the domains of any known boson
families.  Since its observation in 2007, the perplexing complexity of the
invariant mass spectrum and the anomalous nature of the boson masses
made it difficult to find a plausible explanation up to the present
time.

\begin{figure}[h]
\includegraphics[scale=0.50]{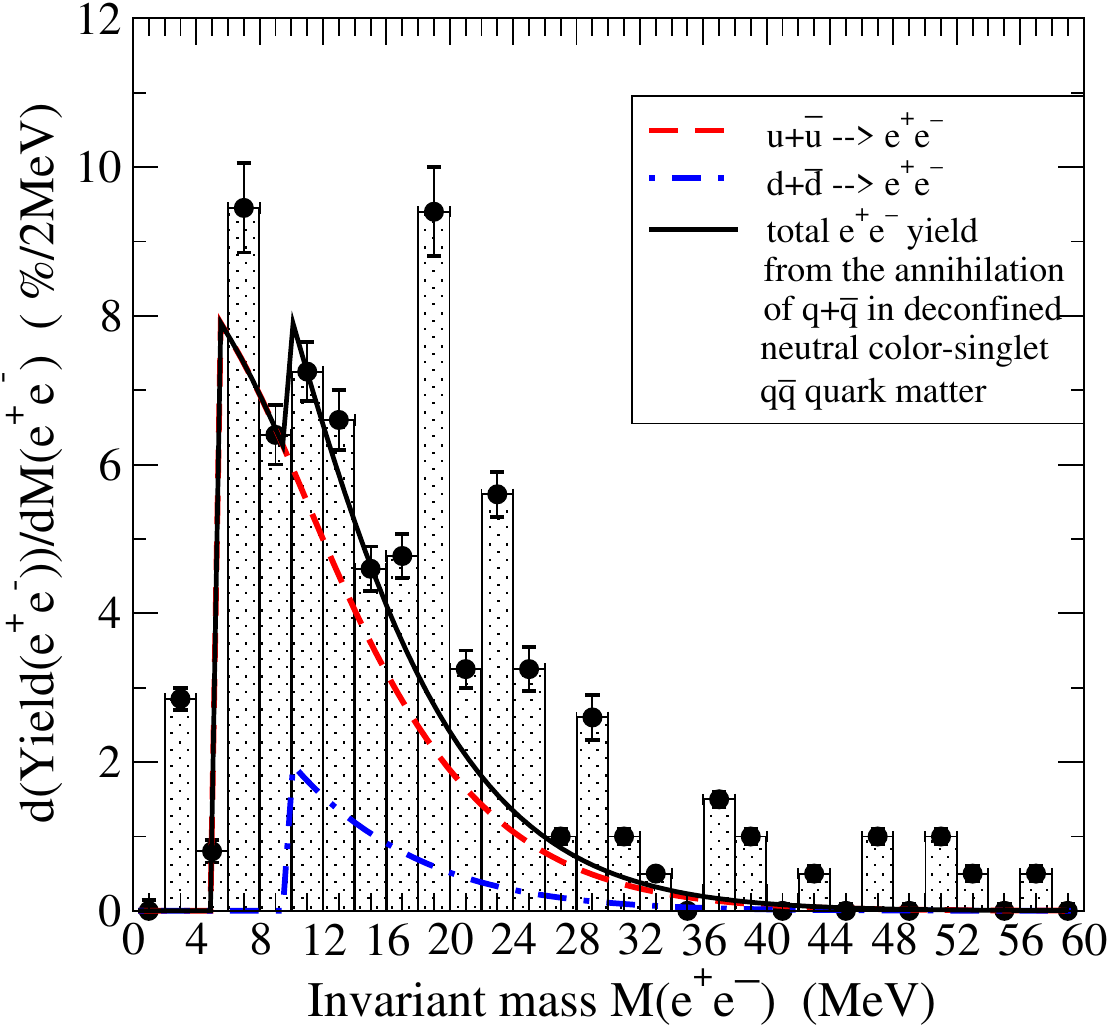} 
\caption{ Shown here is  the experimental
 invariant mass  spectrum of  $e^+e^-$ pairs,
 taken from 
Fig. 4($b$) (Series 1)  of Jain and Singh \cite{Jai07},  
for  $e^+e^-$ pairs that  emerged
  at the distances of $50 \,\mu$m to 200\, $\mu$m away from the
  Pb-emulsion collision vertex in high-energy Pb-emulsion collisions
  at 160 $A$ GeV.  
   The spectrum shows  a broad enhancement at 11$\pm$1 MeV 
  with   a FWHM of of about 10 MeV and additional narrow  resonances within the experimental bin widths of 2 MeV. The resonance energies are listed in Column 6 in Table I.   The solid curve gives  the theoretical model results  of the 
$e^+e^-$ yields 
arising from 
the annihilation of thermal quarks and antiquarks 
  in the QED(U(1))-deconfined neutral color-singlet quark matter, calculated with Eq.\ (\ref{dilep}) for $T=T_c({\rm  QED})=4.75$ MeV, and normalized to the data point at 11$\pm$1 MeV. 
  The experimental resonance energies  are listed  in Table I and will be  compared with  
  theoretical resonance energies from neutral color-singlet quark matter in Section 3.}
\label{fig1}   
\end{figure} 
\begin{table}[h]
\caption {  We 
enumerate  resonances  by the order of resonance energies  in Column 1,  specify  the likely theoretical state in Columns 2 and 3, 
give the equation from which the theoretical resonance energies are calculated in Column 4, and list the theoretical resonance energies in Column 5, 
together with  the  experimental  resonance energies extracted from Ref. \cite{Jai07} reproduced in Figure 1,   and from other references in Column 6.   The theoretical candidate resonance energies  are 
calculated for 
neutral color-singlet $q\bar q$ quark matter
 in  the QED(U(1))-confined 
and QED(U(1))-deconfined phases,
as well as in the QCD(SU(3))-deconfined phase.
The  subscripts ``cnf'' and ``dcnf'' denote  the confined and deconfined phases, respectively.
The  superscripts  ``QED''  labels observed
the neutral color-singlet  $q\bar q$
  quark matter  and 
   the superscript  ``QCD'' labels  
  the neutral color-octet $q\bar q$
  quark matter.  
   Resonance 1 encompasses a spread of experimental energies and 
  might likely be affected by noise backgrounds from  Daltiz $e^+e^-$ decay from pions, which  Jain and Singh    in  \cite{Jai07}  minimized by decay-distance gating.  Resonances with energy greater than 30 MeV may have lower statistical significance because of low numbers of events, they will need further experimental checks and are indicated by parentheses on the resonance energies in Table 1.    Nevertheless, all seven  theoretical candidate resonance energies below 40 MeV match the observed experimental resonance energies of Ref.\ \cite{Jai07} in Table I within the experimental and theoretical energy  uncertainties. 
 }
 \vspace*{0.2cm} 
\begin{tabular}{|c|c|c|c|c|c|}
\hline
Resonance & \multicolumn{3} {c| }{Theoretical Likely Candidate State}&        Theoretical &Experimental  \\ 
    \cline{2-4} 
 Label  & Possible Nature  &   Theoretical  &  Resonance Energy& Resonance &\\ 
                  &       of   State  &  configuration & Calculated  from &        Energy (MeV)  &   Energy  (MeV)           \\ \hline
 1   &            $|u \bar u\rangle_{\rm dcnf}^\qedu  $ &
$ [[q\bar q]^1 \gamma_{ \rm dcnf}^1]^1$ 
&  Eq.\ (\ref{dconf}) for QED&   $5.00\pm 1.8 $   &  \\
   &   $|u \bar u\rangle_{\rm dcnf}^\qcdu  $ &
$ [[q\bar q]^8 g_{ \rm dcnf}^8]^1$
& Eq.\ (\ref{dconf}) for QCD & $4.49\pm 1.6 $  &      \\
&  & & & &  3$\pm 1$\cite{Jai07} \\
&  & & & & $\sim$2\cite{Ana53,deB88} \\
&  & & & & 1.14$\parbox{1cm}{+0.18 \\ --0.07}$,2.1$\pm$0.4\cite{deB88}\\
&  & & & & 1.60$\pm$0.59\cite{Eln88}\\
&  & & & & 1.51$\pm$0.14\cite{Eln07}\\
\hline
  2& $|d \bar d\rangle_{\rm dcnf}^\qedu  $ &
$ [[q\bar q]^1 \gamma_{ \rm dcnf}^1]^1$
& Eq.\ (\ref{dconf}) for QED &   9.90$\pm$3.7 &   \\
  & $|d\bar d\rangle_{\rm dcnf}^\qcdu $  &
$ [[q\bar q]^8 g_{ \rm dcnf}^8]^1$ &   Eq.\ (\ref{dconf}) for QCD & $8.88\pm 3.33 $  
    &         \\
   &   & & & & 7$\pm 1$\cite{Jai07} \\
      &   & & & & $\sim$10\cite{Ana53,deB88}\\  
   &   & & & & 9.2$\pm$1.4\cite{Eln88}\\
 &   & & & & 9.88$\pm2.85$\cite{Eln07}\\
  &  & & & & $\sim$9\cite{deB96,deB97,deB01}\\
 &   & & & & $\sim$9(Fig.\ 15 of Ref.\ \cite{Abr23})\\
\hline
 3 &   $\frac{1}{\sqrt{2}}(|u\bar u \rangle_{\rm cnf}^\qedu\!+|d\bar d\rangle_{\rm cnf}^\qedu )$ & 
$ [[q\bar q]^1 \gamma_{ \rm cnf}^1]^1$
&   Eq.\ (\ref{qed}) for QED& $17.9\pm 1.5 $    & 19$\pm$1\cite{Jai07}   \\
& & & & & 16.94 $\pm 0.12$(stat)$ \pm $0.21(sys)\cite{Kra21}\\
&  &&&&$\sim$17\cite{Abr23}\\
& & & & & 16.66$\pm$0.47(stat)\cite{Tra24}\\
\hline
4 & $|d\bar d\rangle_{\rm cnf}^\qedu $  &     
$ [[q\bar q]^1 \gamma_{ \rm cnf}^1]^1 $
&  Eq.\ (\ref{qed}) for QED&  $22.57\pm 2.2 $  & 23$\pm$1\cite{Jai07}    \\
\hline
 5 &  $|d\bar d\rangle_{\rm dcnf}^\qedu $  $\oplus$ $|d\bar d\rangle_{\rm cnf}^\qedu$   &
 molecular
    & Eq.(\ref{dconf})$\oplus$Eq.(\ref{qed})QED&   $32.47 \pm 4.3$   & 29$\pm$1\cite{Jai07}   \\
\hline
  6 &  $|u\bar u\rangle_{\rm cnf}^\qedu $  & 
   $ [[q\bar q]^1 \gamma_{ \rm cnf}^1]^1$
&   Eq.\ (\ref{qed}) for QED & $33.69\pm 3.4$      & (37$\pm$1)\cite{Jai07}     \\
\hline
  7 &  $\frac{1}{\sqrt{2}}( |u\bar u \rangle_{\rm cnf}^\qedu\!-|d\bar d\rangle_{\rm cnf}^\qedu)$                                   &                                    
   $ [[q\bar q]^1 \gamma_{ \rm cnf}^1]^1$
&   Eq.\ (\ref{qed}) for QED &     $36.4\pm 3.8$       & (39$\pm$1)\cite{Jai07}  \\
  & &&&& $\sim$38\cite{Abr23}\\
\hline
\end{tabular}
\end{table}

It is worth noting that the observation of anomalous neutral bosons
with masses between the photon and pion masses from $e^+e^-$ decays
has a long history, starting at least from 1953 with the ``Bristol''
cosmic ray events examined by Anand \cite{Ana53}.  However, Anand
mis-identified the $e^+e^-$ pairs from anomalous light boson decays as
$e^+e^-$ pairs from Dalitz decays of $\pi^0$ particles, resulting in
the $\pi^0$ lifetime to be wrong by 2 orders of magnitude for about a
decade, as reported in \cite{deB88}.  Re-examination of the data of
the Bristol events by de Boer and van Dantzig gave the anomalous boson
masses to be about 2 and 10 MeV \cite{deB88}.  Through the analysis of
the nucleus-emulsion collision data, De Boer and van Dantzig reported
the observation of neutral bosons at 1.14$\parbox{1cm}{+0.18
  \\ --0.07}$, 2.1$\pm$0.4, and 9.2$\pm$1.4 MeV \cite{deB88}.  El-Nadi
and Badawy reported the observation of anomalous neutral bosons in
nucleus-emulsion collisions at 1.60$\pm$0.59 MeV \cite{Eln88}.
Another study of $^{32}$S-emulsion collisions at 200 A GeV by
El-Nagdy, Abdelsalam, and Badwady \cite{Eln07} reported the
observation of neutral bosons at 1.51$\pm$0.14 MeV and 9.88$\pm2.85$
MeV.

In other high-energy hadron-hadron collisions and $e^+e^-$
annihilations over many decades, many experimental groups observed
anomalous soft photons in the form of excess $e^+e^-$ pairs with low
transverse momenta 
 \cite{Chl84,Bot91,
  Ban93,Bel97,Bel02pi,Bel02,DEL06,DEL08,Per09,DEL10},  
which hinted at the production of parent QED mesons with masses of about 17 and 38 MeV,  as suggested in 
  \cite{Won09,Won10,Won11,Won14,Won20,Won22,Won22a,Won22c,Won23,Won24,Won26}.  In other
nuclear decay experiments, de Boer $et~al.$ observed excess $e^+e^-$
pairs near the invariant mass of 9 MeV \cite{deB97} and confirmed
their observation of the anomaly at 9 MeV in the M1 transition of
$^8$Be at about 10 MeV \cite{deB96,deB01}.  Krasznarhorkay  $et~al.$  \cite{Kra96} reported
the observation of a boson around 12-13 MeV involving $0^-\to 0^+$
transition in $^{16}$O.  
Krasznarhorkay $et~al.$  \cite{Kra16}  later reported
the observation of a boson at 16.70 $\pm$ 0.35 (stat) $\pm$ 0.5(sys)  MeV involving $1^+\to 0^+$
transition in $^{8}$Be.
From the Pb-emulsion spectrum
with many resonances and an enhancement at many different energies, as
we now know in Fig.\ 1, those earlier observations of $e^+e^-$
resonances at many different energies at $\{ \sim$2, $\sim$9, $
\sim$12, $\sim$17$\}$ MeV are likely  the natural consequences of the
complexity of the $e^+e^-$ spectrum as shown in Fig.\ 1. 
The energy locations of these earlier reported neutral boson resonances  
coincide with the energy locations of  the boson  resonances and the peak of the enhancement 
observed by Jain and Singh in Fig.\ 1.
It was not the case that a single  resonance
was mis-identified at many different energies, as it was often assumed
 in the search for a single  elementary particle using $e^+e^-$ pairs.   It was rather the case  that there
were many  $e^+e^-$ resonances occurring at different energies showing up at
many different experiments!  All of these earlier measurements  were  likely  correct
measurements corroborating the resonances in the Pb-emulsion work in Fig.\ 1 of Jain and
Singh \cite{Jai07}.  The presence of the narrow $e^+e^-$ resonance at 19$\pm$1 MeV in high-energy 
Pb-emulsion collisions further cross-confirm the existence of the
hypothetical X17 particle observed at ATOMKI \cite{Kra16,Kra19,Kra21,Kra22,X1722}, DUBNA \cite{Abr23}, and HUS \cite{Tra24} using very different
methods.  It is reasonable to assess that anomalous light neutral
bosons in the 1 to 20 MeV range have been observed by many independent
groups using many different experimental methods (emulsion cosmic ray,
high-energy hadron-hadron collisions, nucleus-emulsion collisions,
nuclear transition spectroscopy, high-energy electron-positron
annihilations, positron-atom collisions) over a span of 70+ years. 
It may also be reasonable to include  the slightly weaker resonance  at 3 $\pm$ 1 MeV in \cite{Jai07} as shown in Fig.\ 1
as a good-signal resonance 
in view of its observation by many other workers \cite{Ana53,deB88,Eln88,Eln07}.
The
persistence of these signals despite repeated skepticism suggests a
real physical phenomenon. It is of great interest to propose a
coherent description of the highly complex structure that characterizes  the perplexing phenomenon.

It should be noted that the invariant mass distribution of Jain and
Singh in Pb-emulsion collisions as presented in Fig.\ 1 was questioned
by de Boer and Fields \cite{deB09,deB11} who performed an independent
re-analysis of the data presented in Fig.\  1($f$) of Ref.\ \cite{Jai07}.
 They
obtained an invariant mass distribution quite different from
Fig.\ 4(b) of Jain and Singh \cite{Jai07}.
However, the re-analysis of de Boer and Fields in \cite{deB09} 
 with only 62 events,
without the important lifetime gating, and without the complete $e^+e^-$ kinematics
may probably be  incomplete 
 because Jain and Singh in
their Fig. 4(b)(series 1)  employed the full sample of 1220 events with
important lifetime gating, full knowledge of $e^+e^-$ kinematics,  and appropriate background subtraction.  In
their follow-up work, de Boer and Fields broadened their scope and 
re-analyzed $e^+e^-$ pair production data from all available emulsion
experiments \cite{deB11}.  They confirmed that something anomalous
existed in emulsion $e^+e^-$ data across multiple experiments and
multiple decades in the mass range of 3 to 20 MeV.  The persistence of
these anomalous signals across multiple experiments, projectile
species, and decades of observations lends credibility to the spectral
results of the Pb-emulsion data of \cite{Jai07}.  It is  therefore
worth  studying  the invariant mass spectrum as reported by Jain and
Singh \cite{Jai07} while acknowledging the high degree of experimental 
uncertainties  and the urgent need for  its
confirmation in future experimental works. The confirmation of the
Pb-emulsion data will become  all the more important because of the
novel and unusual nature of the description we shall propose.
  
Recently, in connection with possible QED(U(1))-confined $q\bar q$ QED
mesons \cite{Won10,Won11,Won14,Won20,Won22,Won22a,Won22c,Won23,Won24,Won26}, it
was pointed out that because quarks and antiquarks belong to the
color-triplet $\bb{3}$ group and the color-antitriplet $\bb{\bar 3}$
group respectively, quarks and antiquarks can combine to form
color-singlet $[q \bar q]^1$ quark matter (CSQM) and color-octet $[q
  \bar q]^8$ quark matter(COQM) \cite{Won26}.  By the
principle of the colorless nature of the observable entities, the
formation of the  lowest-energy $q\bar q$ complexes in color-octet quark matter can arise
only from quarks interacting via the QCD(SU(3)) interaction. 
Similarly, by the same colorless 
principle, the
formation of the  lowest-energy $q\bar q$ complexes in color-singlet  quark matter can arise
only from quarks interacting via the QED(U(1)) interaction. 

The color-singlet quark
matter with two flavors can be further separated into 
neutral and charged color-singlet quark matters (neutral CSQM and charged CSQM). 
The charged color-singlet  $q\bar q$ quark matter 
does not possess stable $q\bar q$ bound states
in the QED(U(1)) interaction because of the repulsion between the quark 
and the antiquark with electric charges of the same sign.  
 Consequently,
charged color-singlet quark matter may not leave 
remarkable signatures of its existence.   Because such a  term appears so often on so many occasions, we shall often conveniently  use the term ``color-singlet quark matter''  to stand implicitly for ``neutral color singlet quark matter'' and add the additional `neutral' or `charged' attribute if ambiguity arises.

In the neutral color-singlet $q\bar q$
quark matter sector, however, a quark and an antiquark interacting only via the
QED interaction may form stable and confined colorless $q\bar q$ QED
mesons  non-perturbatively with masses in the region of many tens of
MeV, arising from the Schwinger confinement mechanism for the open string QED flux-tube
with  the corresponding
attractive linear QED(U(1)) interaction between the quark and the
antiquark
\cite{Sch62,Sch63,Won09,Won10,Won11,Won14,Won20,Won22,Won22a,Won22c,Won23,Won24,Won26}.
The mass of the isoscalar QED meson with $I$=0 and $I_z=0$ was
estimated to be about 17 MeV and the mass of the isovector QED meson
with $I$=1 and $I_z=0$ was estimated to be about 38 MeV
\cite{Won10,Won20}.  
An independent derivation of these masses using quantized Volkov states in (3+1)D QED, without adjusting parameters, has been given by Varro \cite{Var24}, yielding $m_X=17.0$ MeV and $m_E=38.0$ MeV, in agreement with the QED meson predictions.
The experimental observations of the anomalous
soft photons at CERN \cite{Per09,DEL10}, the hypothetical X17 boson
with a mass about 17 MeV at ATOMKI \cite{Kra16,X1722},
DUBNA\cite{Abr23}, and HUS\cite{Tra24}, and another hypothetical E38
boson \cite{Van13} with a mass about 38 MeV at DUBNA\cite{Abr23} hold
promising experimental evidence for the existence of such QED mesons
and possible neutral color-singlet quark matter (NCSQM) at $T=0$
\cite{Won26}.  The prominent presence of a narrow resonance at 19$\pm$1 MeV in
the Pb-emulsion data in Fig.\ 1 \cite{Jai07} provides an independent
confirming evidence for the existence of the hypothetical X17
particle.

It should be mentioned that the X17 particle was speculated to be a
possible carrier of the fifth force \cite{Fen16,Fen20}.  There are
other searches for the X17 particle using different methods
\cite{X1722}.  Recently, the PADME Collaboration searched for the X17
particle by scattering $e^+$ and $e^-$ at around the resonance energy,
using a positron beam to collide with electrons in a diamond target.
They obtained a resonance signal at the expected energy with a
statistical measure of about 2$\sigma$ magnitude \cite{PAD25}, which
is not yet significant enough for a definitive confirmation.  A
measurement with a greater statistical significance is needed.
Another MEG II experiment, originally designed to search for the
$\mu^+ \to e^+ \gamma$ decay, was adapted to investigate the X17 by
studying the $p(^7{\rm Li},e^+e^-)^8$Be reaction, with a proton beam
at an energy at 1.080 MeV colliding with the $^7$Li target, resulting
in the excitation of two different resonances \cite{MEG25}.  The
sensitivity of the measurement has not yet reached the level to observe
a significant signal, and limits on the branching ratios of the two
resonances to X17 were set.

Many experimental and theoretical evidence however strongly support the concept  that  the X17 particle is a
QED meson and one of the stable quanta of the neutral CSQM at $T$=0
\cite{Won10,Won11,Won14,Won20,Won22,Won22a,Won22c,Won23,Won24,Won26}.  
Therefore, the observation of the resonance at 
 19 $\pm$ 1 MeV 
in the  Pb-emulsion spectrum  
and in the experiments at  ATOMKI \cite{Kra16,Kra19,Kra21,Kra22}, DUBNA \cite{Abr23},  and HUS\cite{Tra24} provide 
evidence  for the 
likely existence of neutral color-singlet $q\bar q$ quark matter  at $T$= 0, which in turn, implies the likely
existence of neutral CSQM at $T>0$.  Consequently, there will likely be a
critical temperature $T_c$(QED) below which an assembly of neutral
color-singlet $q\bar q$ quark matter would exist as a QED meson gas,
by analogy with the QCD color-octet quark matter.  Furthermore, for $T > T_c$(QED),
the color-singlet quark matter would likely be QED(U(1))-deconfined
quarks, antiquarks, which may leave an imprint in the Pb-emulsion spectrum.  

We are therefore duly motivated to explore whether the highly complex structure of the $e^+e^-$ spectrum in 
high-energy Pb-emulsion collisions may arise from the neutral
color-singlet $q\bar q$ quark matter in its different phases for the following reasons.  Firstly, we are
keenly encouraged that in nucleus-nucleus central collisions, the
large-mass Pb projectile nucleus colliding with an intermediate-mass
Ag or Br target nucleus, at the high energy of 160 GeV per projectile
nucleon, may provide favorable conditions for the possible production
of deconfined $q\bar q$ quark matter \cite{Won93,Yag05,Vog07}, and the
deconfined quarks and antiquarks in the quark matter can re-combine to form deconfined
color-octet and color-singlet $q \bar q$ quark matters \cite{Won26}.  Secondly,
the masses of the stable  $q \bar q$  quanta of CSQM lie in the same mass region 
as the Pb-emulsion anomaly \cite{Won10,Won20}.  Thirdly, as the
Pb-emulsion anomaly contains a prominent resonance at 19 $\pm$ 1 MeV which
was already identified as a QED meson and a stable quantum of the CSQM
at $T$ = 0 \cite{Won10,Won11,Won14,Won20,Won22,Won22a,Won22c,Won23,Won24,Won26}, the
other parts of the Pb-emulsion anomaly may likely contain other
important elements of the CSQM at $T>0$ or even at $T>T_c$(QED).
Finally, the confinement-to-deconfinement phase transition and the
dynamical evolution of CSQM at different temperatures $T$ contain
rich degrees of freedom and ample complexity which would likely
accommodate the highly complex structure of the Pb-emulsion anomaly.  We are
therefore properly motivated to explore whether the possible
production and evolution of the neutral color-singlet $q\bar q$ quark matter in
high-energy Pb-emulsion collisions may explain the complex Pb-emulsion
anomaly.

To carry out such an exploration, we shall first review the
Pb-emulsion experimental measurements of the anomalous resonances and
the yield enhancement by Jain and Singh \cite{Jai07} in Section 2.
We shall discuss  the degree of statistical significance of the 
Pb-emulsion data prior to comparing them with the proposed theoretical description.
We shall then postulate the production and evolution of the neutral
color-singlet $q\bar q$ quark matter in its two different phases in Section III
to study the spectrum of $e^+e^-$ pairs, when the quark and the
antiquark constituents of the neutral color-singlet quark matter
annihilate into $e^+e^-$ pairs. 
 The comparison of the experimental
$e^+e^-$ invariant mass spectrum with the theoretical $e^+e^-$ invariant mass spectrum from  a
hypothetical neutral color-singlet $q\bar q$ quark matter evolution provides
an initial test whether such neutral color-singlet quark matter may
have been produced in high-energy Pb-AgBr emulsion collisions at CERN
SPS.  We obtain encouraging results as many of comparisons match
favorably in Section 3.  The favorable comparison suggests that both
QED(U(1))-confined and QED(U(1))-deconfined neutral color-singlet $q\bar q$ 
quark matters may have been produced in high-energy Pb-emulsion
collisions at CERN SPS.  Because the possible occurrence of such a
quark matter is extraordinary, the experimental statistics data are
scanty, and the experimental background subtraction
 contains high degrees of uncertainties, it will be
necessary to propose future experiments in Section IV to confirm or to
refute the new findings and to provide additional data for further
explorations.  In Section V, we summarize and conclude the present
discussions. 
A theoretical justification  for plausible  quark confinement in QED(U(1)) in (3+1)D at $T$=0  in the grand unification theory \cite{Geo74,Pat74} of QCD and QED interactions as SU(3)$\times$U(1)/${\mathbb Z}_3$ 
is presented in Appendix A, using concepts proposed earlier by Polyakov \cite{Pol74,Pol87},  t'Hooft \cite{tHo74,tHo75}, Nielsen \cite{Nie73}, Nambu \cite{Nam74}, Mandelstam \cite{Man76}, and Greensite \cite{Gre03,Gre20}.
  
\section{  Observation of Resonances and the Enhancement in High-Energy Pb-emulsion Collisions }
We briefly summarize the main results of Jain and Singh of
Ref.\ \cite{Jai07}.  Jain and Singh exposed photographic emulsion to a
beam of $^{207}$Pb at 160 A GeV at the CERN SPS and selected central
collision events with high shower-particle multiplicity, $N_s$,
between 50 and 800 with a mean multiplicity $\langle N_s \rangle
\approx 270 \pm 21$.  Particle tracks originating at distances away
from the Pb-emulsion collision vertex were followed to look for tracks
of $e^+e^-$ pairs in the very forward direction, within a transverse
region of 100 $\mu$m above and below the primary collision line.
Tracks of produced $e^+e^-$ pairs were identified by two very close
tracks with their characteristic $e^+$ and $e^-$ track ionization
densities, originating from a common $e^+$ and $e^-$ vertex that was
separated from, but properly aligned with, the primary Pb-emulsion
collision vertex.  The alignment with the Pb-emulsion collision vertex
indicates the free travel of a produced unknown neutral parent
particle X in its lifetime, prior to its subsequent decay into the
$e^+e^-$ pair.  The $e^+$ and the $e^-$ energies of the pair were each
measured by their transverse random displacements as a result of their
longitudinal multiple Coulomb scattering through the emulsion medium
\cite{Bar63}.  The distribution of the total energies of the $e^+e^-$
pairs in 1220 events had a peak at the total pair energy of about 300
MeV.  The opening angle between the $e^+$ and the $e^-$ in each pair
was measured by a microscope with an accuracy within $\approx 0.1$
mrad.  The systematic underestimation of the opening angle due to
multiple Coulomb scattering within the emulsion was corrected by the
relative scattering method \cite{Pow59}.  The $e^+$ and $e^-$ energies
and the opening angle of an $e^+e^-$ pair gave the total energy and
the invariant mass $M(e^+e^-)=M(X) $ of the neutral parent particle X
that was produced in the high-energy central $^{207}$Pb-AgBr
collision.  The lifetime of this neutral parent particle X was
obtained from (i) the X particle mean-free travel distance $L_0$
between the primary Pb-emulsion collision vertex and the vertex where
the $e^+$ and the $e^-$ tracks emerged and (ii) the energy and the
invariant mass $M(X)$ of the neutral X particle.  As the invariant
mass $M(X)$ of the unknown neutral parent particle X was the same as
the invariant mass $M(e^+e^-)$ of its decay product $e^+e^-$ pair, the
measurement of the invariant mass spectrum of the $e^+e^-$ pairs,
$d({\rm Yield}(e^+e^-))/dM(e^+e^-)$, was the measurement of invariant
mass spectrum of the unknown parent particle X, $d({\rm
  Yield}(X))/dM(X)$.
 
 The method of photographic emulsion may have the disadvantages of
 limited statistics because of its labor-intensive track measurements
 and the need to scan a vast emulsion area under high magnification
 for track search.  It may however be compensated by the
 advantages of a $4\pi$ acceptance, high detection efficiency (more
 than 90$\%$ for $e^+e^-$ pair detection), superior spatial and
 angular resolutions, photographic track records on the history of
 the observed unknown particle X, the ability to select the range of
 the particle lifetimes by setting gates on the mean-free distance
 $L_0$ between the primary interaction vertex and the $e^+e^-$
 originating vertex in the search, and finally for the present special
 case, the ability to eliminate $e^+e^-$ pairs originating from the
 Dalitz decays of $\pi_0$ particles, which are copiously produced.
 Such an elimination was achieved in Ref. \cite{Jai07} by selecting
 the mean-free distance $L_0$ to exceed 50 $\mu$m, as the mean-free
 path of $\pi_0$ for producing $e^+ e^-$ pairs in emulsion was only a
 few $\mu$m.
 
 From 164 primary events, Jain and Singh selected 1220 $e^+e^-$ pairs
 in the range 50 $\mu$m$ \le L_0 \le 2500 \,\mu$m for examination.
 Only few events were beyond $L_0>$2000\,$\mu$m, and about 30$\%$ of
 events were from $50\, \mu{\rm m} \le L_0 \le 300 \, \mu$m.  The
 invariant masses of the decaying neutral particle, $M(e^+e^-)$,
 spanned from a few MeV up to 90 MeV, while the lifetimes spanned from
 $10^{-15}$\,s to $10^{-12}$\,s \cite{Jai07}.  The transverse momentum
 distribution $d({\rm Yield})/dP_T$ of the produced $e^+e^-$ pair had
 a peak at about $P_T$(peak)$\sim$\,6 MeV and the yield decreased to
 nearly zero at $P_T\sim$ 16 MeV.
 
  After subtracting the background which contributed mainly in the low
  invariant mass region, Jain and Singh examined the invariant mass
  spectrum of the $e^+e^-$ pairs in different maximum ranges of the
  vertex separation distance $L_0$.  They found that the spectrum of
  the $e^+e^-$ pairs depended on the maximum value of the range of
  $L_0$, corresponding to different selections on the range of the
  neutral particle decay lifetimes.
 
 Upon selecting the range of the separation distance $L_0$  between the
 Pb-emulsion collision vertex and the $e^+e^-$ originating vertex
  to be $50\, \mu{\rm m} \le L_0 \le 200\, \mu{\rm m}$, Jain and
 Singh obtained the spectrum as shown in Fig.\ 4(b) (New Series 1)  of \cite{Jai07} reproduced in Fig.\ 1 here.
 As a further measure
 of the $P_T$ values of the resonances, they found that when
 they selected low $P_T$ signals by employing a cut with $P_T \le$\,10
 MeV/c, the yields at the resonances increased, indicating that these
 resonances were associated with neutral particles with low $P_T$
 values below 10 MeV/c.
 
 The invariant mass spectrum in Fig.\ 4(b)(new series 1) of \cite{Jai07} shown here as Fig.\ 1  exhibits two resonances 
 at $7 \pm1$ MeV and $19 \pm 1$ MeV, 
 resting on top of
 an enhancement 
 at $6 \lesssim M(e^+e^-) \lesssim 20$ MeV.
The  resonance  at $7 \pm1$ MeV was also observed by other workers in earlier work \cite{Ana53,deB88,Eln88}.
The 19 $\pm$ 1 resonance was later observed in ATOMKI \cite{Kra16,Kra19,Kra21,Kra22}, DUBNA\cite{Abr23}, and HUS\cite{Tra24}.
There are also weak  local resonance  peaks  with a low number of counts. 
The slightly weaker resonance at 3 $\pm$ 1 may be considered a good-signal resonance 
because it was observed also by many other workers \cite{Ana53,deB88,Eln88,Eln07}.
The energy of the peak of the broad enhancement at 11 $\pm$ 1 MeV coincides with the peak of the 
$e^+e^-$ enhancement  observed by Krasznarhorkay $et~al.$  in \cite{Kra96}.
 It is therefore reasonable to conclude that there have been considerable corroborations of the Pb-emulsion data by  other laboratories.

Because the Pb-emulsion work  in Fig.\ 1 in \cite{Jai07} has low number of event counts, 
questions have often been raised concerning  the statistical  significance of the Pb-emulsion data.
  To answer this question,  
we need first to 
convert the percentage in a bin size of $\Delta M=2$ MeV  at  $M$  to  the numbers of event count in the bin,  $n(M)$.   We  then  need to study the experimental measurement windows, gates, filters, and auxiliary conditions   to come up with  an expected number of background counts.  As raw data of count numbers  are not available after so many intervening years, it suffices to get the best estimates on the number of event counts in these measurements.
We note from Fig.\ 1(a) of \cite{Jai07} that 
the percentage count of 
$e^+e^-$ pairs 
 with the selected vertex separation $L_0$  between 50 micron and 200 micron is approximately   
15\%,  which represents   about 183  $e^+e^-$ pairs
 for a total of  1220 pairs.  Thus, the 100 percentage count in Fig.\ 1 corresponds to  183 event of $e^+e^-$ pairs.    The peak number of count per 2 MeV  bin  at $7\pm 1$ MeV and $19\pm 1$ MeV in the complex structure in Fig. 1 is 17 events.
   The numbers of counts in the complex structure observed by Jain and Singh, even though low in numbers, may  nevertheless provide a high degree of statistical significance as we shall now explain.
   
The counting probability  $P(n,\{s,b\}) $  
at the invariant mass bin $M$  with a bin  width 2 MeV 
in Fig.\ 1  is described appropriately by the Poisson distribution
 in which the probability $P$ for the occurrence of $n(M)$ events  with an expected mean number of signal events $s(M)$  and an  expected mean number of  background events $b(M)$  
 is given by
  \begin{eqnarray}
P(n,\{s,b\}) = \frac{(s+b)^n e^{-(s+b)}}{n! }.
\end{eqnarray}
In 
 the Pb-emulsion data in Fig.\ 1, each of the signal $s(M)$ events  represents  the observation of a pair of   $e^+$ and $e^-$  that originate at a distance 
   of 50 micron to 200 micron  from the collision vertex. 
   The  sum vector momentum of the $e^+e^-$ pair aligns properly with  the Pb-emulsion collision vertex to indicate
   that the pair originates from the decay of a neutral boson produced in Pb-emulsion collisions.   The gating of the separation distance $L_0$ beyond  50 microns provides the filter to  eliminate most, though not all,  of the $e^+e^-$ pairs from pion Dalitz decays. 
 Background $e^+e^-$ pairs $b(M)$ satisfying the selection criteria of windows, gates, filters, and observation conditions  can only be  from the residual  Dalitz pairs and   photons produced in 
 Pb-emulsion collisions and  
 converted  in the emulsion material, with an invariant mass of only a few MeV. Such a  residual  background $e^+e^-$ pairs contribute to the the $e^+e^-$ yield with an $M$-dependent contribution to the invariant mass spectrum, $b(M)$, which   diminishes with increasing invariant mass $M$ beyond a few MeV (see Fig. 4(a) of \cite{Jai07}). The background yield $b(M)$ can be appropriately subtracted as is carried out in Fig. 4(b) of \cite{Jai07}.  
 As a consequence, in the region with invariant mass greater than a few MeV, the diminishing background contributions $b(M)$ 
yields a large statistically significant signal $s(M)$ comprising of resonances resting on top of an enhancement in Fig. 1.

We can investigate the statistical significance of the Pb-emulsion data by calculating  the Poisson
$p$-value and its equivalent Gaussian standard deviation
based on $s(M)$ and $b(M)$.
The Poisson $p$-value is defined as 
\begin{eqnarray}
p{\rm -value}=\sum_{k=n}^\infty \frac{b^k e^{-b}}{k!}.
\end{eqnarray}
We can define statistical significance as the $Z(\sigma)$ number of standard deviations
that a Gaussian variable would fluctuate in one direction
to give the same Poisson $p$-value. 

For the broad enhancement from 5 to 21 MeV with a maximum at $11\pm1$
MeV in Fig. 1, the total count number in Fig. 4(a) of \cite{Jai07} is
about 128 and the background count is about 43, leading to an
aggregate significance of $Z(\sigma)=10.4\sigma$. Removing the
resonances at $7\pm 1$ MeV and $19\pm 1$ MeV does not change the
aggregate significance significantly. In either way, the broad
enhancement itself remains highly significant even after removing the
known resonance contributions.

For the resonance  at $19 \pm 1$ MeV a stand-alone resonance with $s(M )$=17 and $b(M ) $$<$0.5, in
Fig. 4(a) of \cite{Jai07}, we get  $p$-value less than  $10^{-16}$, equivalent to a Gaussian significance of greater than 8$\sigma$.
There however may be two sources: $s_1(M)$ from the broad
enhancement, and $s_2(M)$ from an additional narrow resonance within
the experimental bin width of 2 MeV. To obtain the statistical
significance of the second source $s_2(M)$, we use the fit to the
first source $s_1(M)$ as the background for $s_2(M)$. For the bin at
$19\pm 1$ MeV, we have $s_1(M)\sim 5$ from the solid curve fit to the
broad enhancement, so $s_2(M) = 17 - 5 = 12$ counts, and the
background for the second source is $b_2(M) = s_1(M) \sim 5$. The
$p$-value is obtained from the Poisson tail evaluated at the total
observed count of 17 against this background, $p = P(N \ge 17 \mid
b_2 = 5)$, giving $p = 1.4\times 10^{-5}$ for the second source
$s_2(M)$, with the corresponding statistical significance
$4.20\sigma$. A 30\% error in $b_2(M)$ up to $b_2(M)=6.5$, so as to
include the two (higher) next-neighboring points of the resonance,
will reduce the statistical significance to $3.42\sigma$. One can
therefore consider the $19\pm 1$ MeV resonance above the thermal
quark yield background to have a statistical significance from 3.4 to
4.2$\sigma$ if it is a resonance resting on top of the thermal quark
annihilation background enhancement.

For the resonance  at $7 \pm 1$ MeV, the signal count after subtracting  the noise background  is $s(M)=$17.  
If we describe this resonance as  a stand-alone resonance, we can estimate the noise background from the two next-neighbor points with  $b(M)=10$ in Fig. 4(a) of \cite{Jai07}.  We obatain $p$-value=6.42$\times 10^{-6}$ or  a statistical significance of $4.36 \sigma$.  

An alternative description is that the yields in this mass bin  at $7\pm 1$ MeV  come predominantly from
the annihilation of thermal quarks, as determined by the theoretical fit to the broad enhancement carried out in section 3.3 and show in Fig 1.  In that case,  the fit from the annihilation of thermal quarks gives the signal yield   $s(M)\sim 13$, which is close to the total yield of 17 counts, leave only 4 counts  for a  possible low-yield resonance at $7\pm 1$ MeV.    As we discuss in Eq.\ (\ref{eq1}) in Section 3.4, such a low-yield resonance may be expected, as the resonance yield for this resonance would be only approximately 1/4 
of the resonance yield at $3\pm 1$ MeV, if the resonances at $3\pm 1$ and $7\pm 1$ are Columb bound states of the $u \bar u$ and $d \bar d$ states of the deconfined neutral color-singlet quark matter. 

It is therefore reasonable to conclude from the above discussions that there are considerable corroborations 
from many other laboratories on  part of the Pb-emulsion experimental data.  Furthermore,  the resonance peaks and the broad enhancement have different   degrees of statistical significance  for different parts of the highly complex  spectrum.
 
We shall present  a coherent theoretical description for  the  highly complex  $e^+e^-$ invariant mass spectrum in high-energy Pb-emulsion collisions obtained in Ref.\  \cite{Jai07}.

\section{Theoretical  States  of Color-Singlet and Color-Octet $q\bar q$ Quark Matters}

\subsection{Evolution of the dense matter produced in high-energy Pb-emulsion central Collisions}

 As we have described in the Introduction, we have been well motivated to explore whether the production and the
 evolution of the neutral color-singlet  $q\bar q$   quark matter (CSQM)  in high-energy
 Pb-emulsion collisions may explain the Pb-emulsion data.   We wish
 to study a hypothetical production and evolution of CSQM and COQM in
 order to predict a theoretical $e^+e^-$ spectrum in $q\bar q$
 annihilation, to compare with experimental Pb-emulsion data.

The dynamical evolution of the quark matter after production will go
through different phases.  Each phase of the different $q\bar q$ quark
matters will produced different sources of $e^+e^-$ pairs.  The
perplexing nature of the highly complex structure  of the spectrum,  with many anomalous
resonances on top of  a broad yield enhancement,  necessitates the exploration into
all possibilities, including some unfamiliar and unusual ones.

In our search for the states of different $q\bar q$ quark matters in
its different phases, we hold the principle that nature will avail
itself of all opportunities that are not explicitly excluded.  Thus,
all possible entities interacting in both confined and deconfined
QCD(SU(3)) and QED(U(1)) interactions, leading to states in the
anomalous region below 50 MeV will be considered.  The suggested
sources of some of the $e^+e^-$ pairs can only be considered
tentative, pending future confirmation of the highly perplexing experimental
data.
 
We envisage a central Pb-AgBr collision at 160 A GeV which produces
hadron matter with a very high energy density in a very small region
of space, in the central rapidity region of momentum, in a very short
interval of time.  In such an extreme condition, the QCD hadron matter
may likely thermalize to a temperature $T$ greater than the critical
QCD $q\bar q$ confinement-to-deconfinement temperature $T_c$(QCD) to
become deconfined COQM as a quark gluon plasma
\cite{Won93,Yag05,Vog07}.  Deconfined quarks and deconfined antiquarks
in the quark gluon plasma will combine to form the deconfined
color-octet quark matter and the deconfined color-singlet quark matter
interacting via the color-octet QCD((SU(3)) interaction and the
color-singlet QED(U(1)) interaction with different coupling constants
respectively \cite{Won10,Won20,Won26}.
 
As the QCD(SU(3))-deconfined and QED(U(1))-deconfined $q\bar q$ quark
matter cools to the transition temperature, $T$=$T_c$(QCD), the
color-octet quark matter makes a transition from the quark-gluon
plasma phase with deconfined quarks to a hadron gas phase with
QCD(SU(3))-confined QCD mesons.  When the temperature $T$ decreases
further to $T_c$(QCD)$ > T > T_c$(QED), the QCD(SU(3))-deconfined
quarks in the color-octet quark matter in the quark gluon plasma
remain QCD(SU(3))-confined as a QCD meson gas, whereas the quarks in
the neutral color-singlet quark matter remain QED(U(1))-deconfined.

\subsection{Deconfined quarks and antiquarks are subject to U(3)-confinement and a change of interactions}

It should be clarified at the outset that the QCD(SU(3))-deconfinement
and QED(U(1))-deconfinement of quarks and antiquarks do not
necessarily imply the presence of spatially isolated observable quarks
and antiquarks.  Quarks carry color and they in fact interact via a
broken U(3) interaction which can be broken up into the QED(U(1))
interaction and the QCD(SU(3)) interaction, possessing different
coupling constants \cite{Won10}.  The principle of colorless nature of
physical observable entities is a principle of U(3)-confinement.  That
is, colored particles such as quarks interacting in an U(3)
interaction, are bounded spatially by a U(3)-boundary such that no
single U(3) particle can be spatially isolated.  Hence, the
QCD(SU(3))-deconfined and the QED(U(1))-deconfined quarks and
antiquarks remain U(3)-confined in a greater U(3)-boundary and they
continue to be subject to a boundary effect \cite{Won93a} after their
QED(SU(3)) and QED(U(1)) deconfinement.  The deconfined quarks and
antiquarks do not possess truly continuum states that extend
asymptotically in space as they remain bounded by the U(3)-boundary as
quasi-continuum states.  The U(3)-boundaries for the CSQM and the COQM
need not be the same as their underlying U(1) and SU(3) interactions
of the U(3) interaction have different coupling constants.
   
 In the deconfined state, the annihilation of such a ``thermal'' quark
 with its antiquark to decay into an $e^+e^-$ pair will take place to
 yield a quasi-continuum set of $e^+e^-$ pair spectrum.  We shall look
 at the emulsion data to see if such quasi-continuum states of the
 deconfined thermal $q \bar q$ pairs may show up in the spectrum of
 the $e^+e^-$ pairs.

  At the temperature of $T_c$(QED) for QED(U(1))-confinement-to
  deconfinement phase transition in a neutral color-singlet quark
  matter, the change of the confinement behavior is in fact a change
  of the mutual interaction between the quark and the antiquark.  In
  the case of the QCD-confinement to QCD-deconfinement phase
  transition, the change of the interaction at the phase transition can
  be approximately considered as a change of a
  linear-plus-color-Coulomb to just a color-Coulomb type interaction
  without the linear interaction and the dissolution of the flux tube
  between quarks and antiquarks.  We shall discuss the new
  interactions and the consequences in Section D.

\subsection{ The yield enhancement in the $e^+e^-$ Spectrum as a Signature for Deconfined Quarks}

In the deconfined phase of quarks and antiquarks, there are two
distinct deconfinement signatures in the $e^+e^-$ spectrum.  The first
deconfinement signature comes from the spectrum of $e^+e^-$ pairs
which arise from the annihilation of deconfined quarks and antiquarks
in quasi-continuum $q\bar q$ states, and such $e^+e^-$ pairs will lead
to a relatively broad yield enhancement extending over a broad region
of $e^+e^-$ invariant masses whose width will be of the order of the
temperature at the moment of annihilation.

As a system undergoing a phase transition spends the greatest
proportion of time at the phase transition, the annihilation of the
quark and the antiquark will likely take place at the transition
temperature, $T_c$(QCD) for QCD(SU(3)) phase transition of the quark
gluon plasma, and $T_c$(QED) for the QED(U(1)) phase transition of the
quark photon plasma.  The measurement of the shape of the $e^+e^-$
spectrum will provide direct information on the critical phase
transition temperature, $T_c$(QCD) or $T_c$(QED), as the case may be.
The threshold of the yield enhancement depends on the masses of the
annihilating deconfined quarks.

The second signature has to do with a change of the mutual interaction
between the quark and the antiquark after deconfinement.  These
changes will lead to new bound states, showing up as new resonances in the
$e^+e^-$ invariant mass spectrum at specific bound state energies.

In the first signature in the deconfined phase, there are quarks and
antiquarks in a quasi-continuum state within a U(3)-boundary.  We
shall however approximate such a quasi-continuum state as a continuum
state, for ease of theoretical calculations.  The quark and an
antiquark in the continuum can annihilate as depicted in Fig.\ 2(a).
Their annihilation through such a diagram as Fig.\ 2(a) provides
information on the momentum distribution of quarks an antiquarks in
the deconfined phase.

\begin{figure}[h]
\includegraphics[scale=0.80]{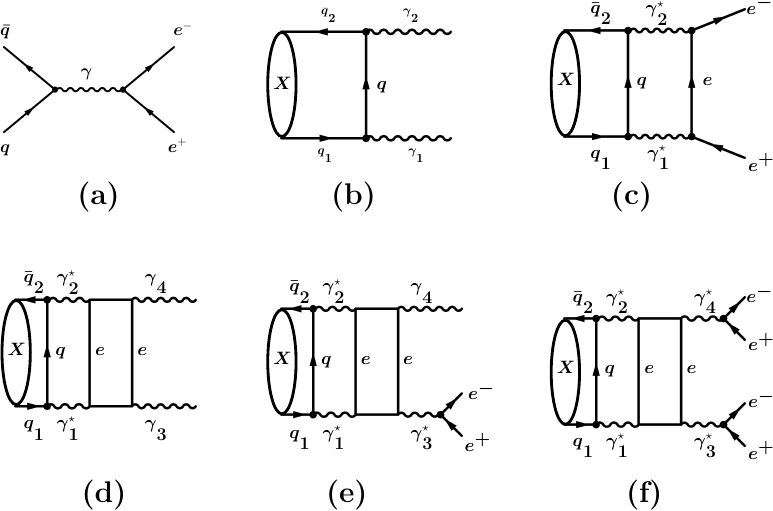} 
\vspace{0.8cm}
\caption{ Decay diagrams of different orders for the decay of the
  $q\bar q$ state $X$ in a $q\bar q$ NCSQM or COQM into real or
  virtual photon pairs or an $e^+e^-$ pair.  The lifetime for $e^+e^-$
  decay through diagrams ($a$) or $(b)$ is of order $1/\alpha^2$. The
  lifetimes for decay through diagram $(c)$ is of order $1/\alpha^4$,
  and through $(d)$, $(e)$ and $(f)$ into an real and virtual photons
  are of order $1/\alpha^6$.  }
\label{fey}   
\end{figure} 
As it has been examined previously, the invariant mass distribution of
the $e^+e^-$ pair distribution is given by Eq. (6) on page 314 of
\cite{Won93} or equivalently in terms of Bessel function $K_1(M/T)$ in
\cite{Kaj86}
\begin{eqnarray}
\frac{dN_{q \bar q \to l^+l^-}}{dM^2 d^4x} = N_c N_s^2 \sum_{f=1}^{N_f} \frac{\sigma_f (M)}{2(2\pi)^4}  M^2\left ( 1-\frac{4m_q^2}{M^2} \right )^{\frac{1}{2}} T \sqrt{ \frac{\pi M T}{2} }e^{-M/T},  
\label{dilep}
\end{eqnarray}
where the number of colors $N_c$ is 3, the number of spins $N_s$ of
quarks is 2, and the annihilation cross section $\sigma_f(M)$ is
\begin{eqnarray}
\sigma_f (M)=\frac{4\pi}{3} \frac{Q_f^2(\alpha^\qedu)^2}{M^2} \left ( 1-\frac{4m_q^2}{M^2 }  \right )^ {-\frac{1}{2} }
 \left ( 1-\frac{4m_l^2}{M^2 }  \right )^ {\frac{1}{2} } \left( 1+2\frac{m_q^2+m_l^2}{M^2} +4 \frac{m_q^2 m_l^2}{M^4 }\right ),
 \label{sigma}
\end{eqnarray}
where $Q_u=2/3$ and  $Q_d=-1/3$ are the charge numbers of the quark with flavor $f$, and $\alpha^\qedu=1/137$ is the QED fine-structure constant.

In the evolution of the dense matter after their production, the
system spend their dominant proportion of their times at the phase
transition temperature, and the annihilation of the quark and
antiquark takes place predominantly at the phase transition
temperatures.  We can estimate the magnitude of $T_c$(QED) from the
physical condition that, at the critical phase transition temperature
$T_c$(QED), the lowest mass QED meson in the confined NCSQM is in a
transitional equilibrium with its deconfined quark and antiquark
constituents.  Thus, the mass of the lowest energy confined QED boson
will be equal to the sum of the energy of its deconfined constituents
each of which has the kinetic energy of the critical temperature
$T_c$(QED) plus the rest mass energy of the deconfined quark and
antiquark.

We hypothetically identify X17 as the lowest-energy QED(U(1))-confined
isoscalar meson which has the wave function $\frac{1}{\sqrt{2}}(|u\bar
u \rangle_{\rm cnf}^\qedu\!+|d\bar d\rangle_{\rm cnf}^\qedu )$.  The
mass of the lowest-energy confined QED meson in the confined phase
before deconfinement is just $M$(X17).  After deconfinement the total
rest mass of the two deconfined constituents is the average, $\frac{1}{2}[
  (m_u+m_{\bar u})$+$(m_d+m_{\bar d}) ]$, and each of the two
deconfined constituents acquires a kinetic energy of the critical
temperature $T_c$(QED) in a longitudinal break up,
\begin{eqnarray}
M(X17) \sim \frac{1}{2} [(m_u+m_{\bar u})   + (m_d +m_{\bar d})] + T_c({\rm QED})+ T_c({\rm QED})
\end{eqnarray}
which gives
\begin{eqnarray}
T_c({\rm QED}) \sim \frac{1}{2}[  M(X17) - m_u - m_d ].
\label{tc}
\end{eqnarray}
If we take $M$(X17) to be 16.94 $\pm 0.12(stat) \pm 0.21 (sys)$ MeV from the ATOMKI measurement \cite{Kra21} and the quark masses 
$m_u=2.5 \pm 0.90$ MeV and $m_d=4.95 \pm 1.85$ MeV from PDG \cite{PDG18}, we get 
\begin{eqnarray}
T_c({\rm QED}) \sim  4.75 \pm 1.04 ~{\rm MeV}. 
\label{eq5}
\end{eqnarray}
Thus, $T_c$(QED) is much lower than the corresponding theoretical
QCD(SU(3)) deconfinement temperatures $T_c$(QCD) of about 150-160 MeV
for $\mu_B=0$ \cite{Bor24}. The possible classification of quark
matter in color space, temperature $T$, and phases is depicted in
Fig. 1 of Ref. \cite{Won26}.

In the deconfined neutral color-singlet quark matter, the annihilation
of the quarks and antiquarks into $e^+e^-$ pairs give rise to the
$e^+e^-$ yield enhancements estimated from Eq.\ (\ref{eq5}) and shown
as curves in Fig.\ (1) where we have used $T=T_c({\rm QED})=4.75$ MeV,
and $m_l=m_e=0.511$ MeV.  The sum $(e^+e^-)$ invariant mass yield
enhancement of $d({\rm Yield})/dM(e^+e^-)$ is shown as the solid curve
in Fig.1, normalized to the  peak of the experimental yield enhancement 
at $M(X)$ at 11 $\pm$ 1 MeV.
The contributions from the QED(U(1))-deconfined $d$ and $u$ quarks are
given in the dashed and the dashed-dot  curves, respectively.
The $u$ quark contribution is about 4 times of that from the $d$ quark
because $|Q_u| = 2 |Q_d|$.  The temperature $T=4.75$ MeV is the
estimate of $T_c$(QED) from the X17 mass and the quark masses in
Eq.\ (\ref{tc}).  The approximate agreement of the shape of
experimental enhancement with the shape of the theoretical total yield
enhancement suggests the possibility that QED(U(1))-deconfined neutral
color-singlet quark matter may have been produced in high energy
Pb-emulsion collisions in Ref.\ \cite{Jai07}.

\begin{figure}[h]
\includegraphics[scale=0.40]{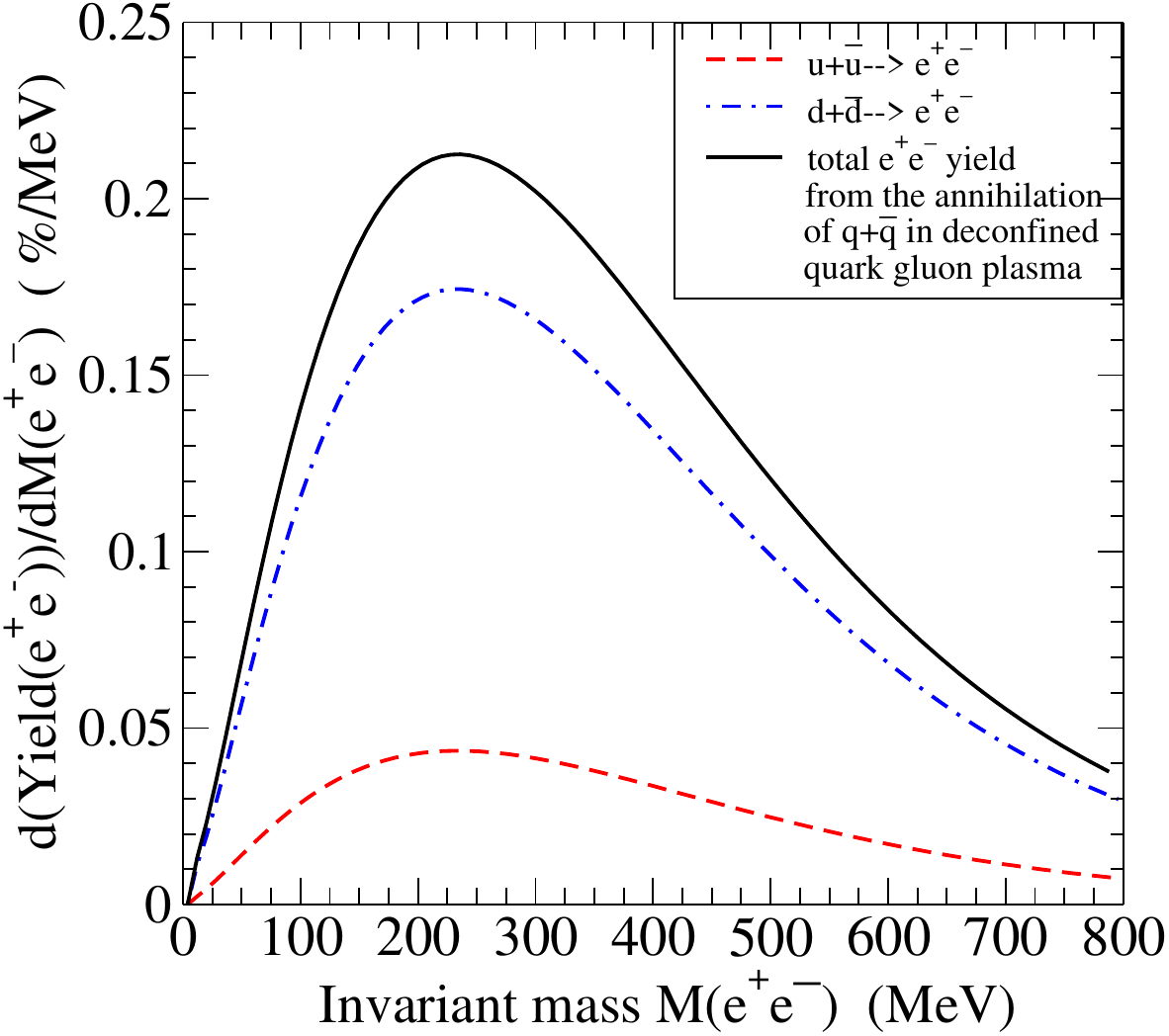} 
\vspace{0.8cm}
\caption{ Theoretical distribution of the total yield of $e^+e^-$
  pairs as a function of invariant mass of the $e^+e^-$ pair arising
  from the annihilation of the light quarks and antiquarks in
  deconfined quark gluon plasma at $T=T_C$(QCD)$=156$ MeV, normalized
  to 100 percent. }
\label{ncoqm}   
\end{figure} 

For completeness, we can examine the $e^+e^-$ spectrum from the
related QCD(SU(3)-deconfined color-octet quark matter, the quark-gluon
plasma.  There should be a similar $e^+e^-$ yield enhancement as a
function the $e^+e^-$ invariant mass arising from the annihilation of
quarks and antiquarks in QCD(SU(3))-deconfined quark gluon plasma
interacting via the color SU(3) interaction.  As only quarks of the
same flavor can annihilate, the $(e^+e^-)$ invariant mass distribution
from the annihilation of deconfined quarks will probe the quark and
antiquark distributions of the neutral color-octet quark matter
(COQM).  It is of interest to make a prediction on the shape and the
peak of the $e^+e^-$ invariant mass yield enhancement.  For such a
calculation, we can combine Eqs.\ (\ref{dilep}) and (\ref{sigma}) to
get
 \begin{eqnarray}
\frac{dN_{q\bar q\to l^+l^-}}{dM^2 d^4x} = N_c N_s^2 \sum_{f=1}^{N_f} \frac{(Q_f\alpha)^2}{3(2\pi)^3}  \left ( 1-\frac{4m_l^2}{M^2} \right )^{\frac{1}{2}} 
\left( 1+2\frac{m_q^2+m_l^2}{M^2} +4 \frac{m_q^2 m_l^2}{M^4 }\right )
T \sqrt{ \frac{\pi M T}{2} }e^{-M/T}.  
\label{eq6}
\end{eqnarray}
For the QCD(SU(3))-deconfined case, we have $\{m_q, m_l \} \ll T$ and the quark masses can be neglected, and we get
\begin{eqnarray}
\frac{dN_{q \bar q \to l^+l^-}}{dM d^4x}  \propto  M^{3/2} e^{-M/T}. 
\end{eqnarray}
The peak of the $e^+e^-$ yield distribution is located at
\begin{eqnarray}
M(e^+e^-)|_{\rm peak} \sim \frac{3}{2} T.
\end{eqnarray}
For a confinement-to-deconfinement phase transition temperature of
$T=T_c(QCD)=156$ MeV for COQM, the peak of the yield enhancement is
located at $T=234$ MeV, and the distribution has a FWHM of 460 MeV in
Fig.\ \ref{ncoqm}.  The $e^+e^-$ pair yield is monotonically
increasing as an increasing function of the $e^+e^-$ invariant mass in
the region below 50 MeV as shown in Fig.\ \ref{ncoqm}.
   
It is instructive to compare and contrast the shapes and the peak  locations of the theoretical 
$dN_{q\bar q \to e^+e^-}/dM(e^+e^-) |_{\rm (CSQM)}^{\rm dcnf}$ 
in the annihilation of thermal quarks and antiquarks in deconfined color-singlet  quark matter
CSQM in Fig. 1,
 with the theoretical  
$dN_{q\bar q \to e^+e^-}/dM(e^+e^-)|_{\rm (COQM)}^{\rm dcnf}$ 
in the annihilation of thermal quarks and antiquarks in deconfined color-octet quark matter
(COQM)(i.e. quark gluon plasma)
in Fig. 3.
 Their shapes are quite different and the peak location relative to the phase transition temperature
 $T_c$  are quite different.
For the CSQM in Fig. 1, $T_c(QED)$ is 4.75 MeV which  is of the same order as the quark masses of $m_u=2.5$ MeV and $m_d=4.95$ MeV, and the threshold effects show up very clearly in Fig 1, with sharp increase in the yield just at the quark mass threshold, $2m_q$.  The peak of the total yield is close to the mass threshold of the $2m_u$ from $u$  quark pairs, as the $u$ quarks contribute 4 times the cross section compared to the $d$  quarks.   The shape of the yield is sensitive to  the quark masses which may be utilized as a direct experimental measurement of the up and down quark masses in deconfined color-singlet quark matter.
In contrast, for the COQM, $T_c(QED) \gg m_q$, the yield is insensitive to the quark masses, and the peak of the distribution is located at $3T/2$.

\subsection{New deconfined $q\bar q$ States as a Signature for Deconfined Quarks}

We mentioned that when a $q\bar q$ quark matter makes a phase
transition from the confined state to the deconfined state, the mutual
interaction between a quark and its partner antiquark changes.  As
consequence, there can be new $q\bar q$ bound states in the new
interaction in the deconfined quark matter.  The quark and the
antiquark of the bound state can annihilate and decay into a $e^+e^-$
pair, as in Fig.\ 2($c$) and show up as resonances in the $e^+e^-$
invariant mass spectrum in Fig.\ \ref{fig1}.  These $e^+e^-$
resonances as decay products of new $q\bar q$ bound states can be
the signatures for the deconfined $q\bar q$ quark matter.

We can understand the change of the interaction and the occurrence of
the new $q\bar q$ bound states in the following way.  In the confined
$q\bar q$ matter prior to the deconfinement phase transition,
confinement occurs as the color-Coulomb lines of force for
QCD(SU(3))-confinement, or the electric-Coulomb lines of force for the
Schwinger-type QED(U(1))-confinement, are bundled together between the
quark and the antiquark in the form of a confined flux tube.  By Gauss
law of electric fluxes, the bundled Coulomb lines of force in the flux
tube lead to a linear interaction between the quark and the antiquark
and the subsequent QCD(SU(3))-confinement or the
QED(U(1))-confinement.  The interaction between the quark and the
antiquark is then a linear-plus-Coulomb interaction as in the
Cornell-type phenomenological potential for QCD mesons, or the
L\"usher-type linear-plus-Coulomb expansion of the string interaction
for QCD and QED mesons
\cite{Eic75,Lus80,Lus81,Bar92,Won00,Won01,Won22c}.

  As temperature increases, the color-Coulomb or the electric-Coulomb
  lines of forces between the quark and the antiquark begin to
  de-bundle from the flux tube and will spread out to an extended
  spatial region in three dimensional space.  As temperature increases
  to the critical deconfinement temperature, the de-bundling of the
  color-Coulomb or the electric-Coulomb lines of force of the flux
  tube will be complete with the disappearance of the flux tube, the
  disappearance of the linear interaction, and the occurrence of quark
  deconfinement
   
   After the deconfinement transition takes place, the color-Coulomb
   or electric-Coulomb lines of forces extend to the full
   three-dimensional space up to the U(3) boundary.  That is, the new
   interaction between a quark and an antiquark in the deconfined
   quark matter will be the color-Coulomb interaction for the
   QCD(SU(3))-deconfined color-octet quark matter and the
   electric-Coulomb interaction for the QED(U(1))-deconfined
   color-singlet quark matter.  The interaction potential between a
   quark $q$ and its antiquark partner $\bar q$ at a distance $r$ will
   be
  \begin{eqnarray}
V_{q\bar q}(r)=\alpha^\lambda \frac{Q_q^\lambda Q_{\bar q} ^\lambda }{r}
\end{eqnarray}
where $\lambda$ is the label ``QCD'' or ``QED'' for ``QCD(SU(3))
interaction'' or ``QED(U(1)) interaction'', respectively,
$Q_q^\lambda$ is the effective quark color charge or the electric
charge number for QCD or QED respectively, and $Q_{\bar q}^\lambda
=-Q_q^\lambda$.  The forces are attractive between a quark and its
antiquark partner in the deconfined quark matter, leading to Coulomb
bound states in the deconfined quark matter.  The mass formula for the
$|q\bar q\rangle_{\rm dcnf}^\lambda$ bound state in the Coulomb
interaction is
\begin{eqnarray}
M( |q\bar q\rangle_{\rm dcnf}^\lambda)
 = 2m_{q} -\frac{1}{2}\mu_q\frac{  (C_F^\lambda Q_q^\lambda \alpha^\lambda )^2}{n^2}
 = 2m_{q} -  \frac{ m_q (C_F^\lambda Q_q^\lambda \alpha^\lambda )^2}{4n^2}, 
\label{dconf}
\end{eqnarray}
where  the color factor $C_F=1$ for QED, and $C_F=4/3$ for QCD, the subscript label ``dcnf'' stands for ``in deconfined quark matter'',   and $n$ is the principle quantum number.

The Coulomb interaction will be screened by the presence of other
charges in the medium.  The screening will modify the Coulomb
interaction to turn it into a Yukawa-type interaction with a finite
range.  
The Coulomb bound state energy obtained by Eq.\ (\ref{dconf})
provides a useful benchmark against which the degree of charge
screening can be assessed as the cross section will be modified accordingly by the 
initial state interaction as discussed in \cite{Won96a,Won96b}.

It should be emphasized that the Coulomb binding energies of Eq. (12) are extremely small,  of order 10 eV to 1 MeV,  compared with the transition temperature $T_c(QED) \sim   4.75$ MeV at which the deconfined quarks and antiquarks are produced. One may therefore ask whether these states survive as genuine discrete poles in a thermal medium, or whether what is being observed is more properly described as a Coulomb (Gamow-Sommerfeld) enhancement of the $q\bar q\to  e^+e^- $ cross section in the immediate vicinity of the $M = 2m_q$ threshold, arising from the residual attractive final-state interaction between the outgoing quark and antiquark. This distinction was previously examined for the analogous case of quark-antiquark final-state Coulomb and color-Coulomb interactions in a screened deconfined medium \cite{Won96a,Won96b}, where it was shown that such interactions modify the near-threshold 
cross section even when a true bound state does not survive the screening length of the plasma. The general phenomenon --  an attractive Coulomb-type final-state interaction enhancing a pair-production cross section at threshold -- is a well-established effect, first derived by Sakharov for $e^+e^-$ pair production \cite{Sak48,Fad95,Sti95}, and confirmed experimentally in $e^+e^- \to  W^+W^-$ production at LEP2, where the measured threshold cross section requires inclusion of the QED Coulomb correction at the several-percent level to be consistent with the independently measured $W $ mass. Whether the resonances identified in Table I as $|u\bar u_{\rm dcnf}\rangle$  and $|d\bar d_{\rm dcnf}\rangle$  correspond to discrete $n=1$ Coulomb bound states or to Gamow-Sommerfeld-enhanced threshold structures in the continuum, the underlying physical origin --  an attractive residual interaction between the deconfined but still $U(3)/{\mathbb Z}_3$-bounded quark and antiquark --  and the resulting near-threshold enhancement of the $e^+e^-$ yield, are the same.

We list the masses of the $q\bar q$ bound states for $n=1$, $|u \bar
u\rangle_{\rm dcnf}^{\{\qcdu, \qedu\}} $, and $|d \bar d\rangle_{\rm
  dcnf}^{\{\qcdu, \qedu\}} $, color-Coulomb and electric-Coulomb
states in Table I, with $\alpha^\qedu=1/137$ and $\alpha^\qcdu=0.68 $
\cite{Won20}.  As the QED fine structure constant $\alpha^\qedu$ is
much smaller than unity, the Coulomb binding energy of quark and
antiquark is small compared to the quark mass, we have the Coulomb
bound state energies of $M( |q\bar q\rangle_{\rm dcnf}^\qedu)\sim 2
m_q$, as listed in Table 1.  In the presence of two types of
interactions, we label symbolically the confining Schwinger-type QED
interaction as the exchange of a confining photon by $\gamma_{\rm
  cnf}$ and a deconfining photon by $\gamma_{\rm dcnf}$, respectively
as shown in Table 1.  Similarly, we label the QCD confining
interaction as the exchange of a confining gluon by $g_{\rm cnf}$ and
a deconfining photon by $g_{\rm dcnf}$, respectively in Table 1.
 
 The  resonance observed at  about 3 $\pm$ 1 MeV in
 Ref.\ \cite{Jai07,Ana53,deB88,Eln88,Eln07}  may 
 be interpreted as one or both of the deconfined $|u \bar u\rangle_{\rm dcnf}^{\{\rm
   QED,QCD\}}$ states, while the resonance observed at about 7 $\pm$ 1  MeV in \cite{Jai07,Ana53,deB88,Eln88,Eln07,deB97,deB96,deB01,Abr23} may be interpreted as one or both of the 
   deconfined $|d  \bar d\rangle_{\rm dcnf}^{\{\rm QED,QCD\}}$ states.
 At the present stage of our measurement, the fine structure of these
 QCD and QED deconfined bound states cannot yet be resolved.

While the above comparison of the resonance energies is one aspect of the check
on the proposed mechanism,
the magnitudes of the resonance yields  can also be a confirming check.
Specifically, we can use  Eq.\ (\ref{dilep}) to double-check on the validity of the resonance description that the $3\pm 1$ resonance and $7\pm 1$ resonance  are indeed  $|u\bar u\rangle _{\rm dcnf}^{\rm res}$ and 
 $|d\bar d \rangle _{\rm dcnf}^{\rm res}$, respectively because the production cross section is proportional to the electric charge number, their cross section yields would differ by a factor close to 4.   Such big factor should show up in the experimental data.
 
The generalization of Eq.\ (\ref{dilep}) for the
$e^+e^-$ yield in  the production of a $|q\bar q\rangle_{\rm dcnf}^{\rm res} $ resonance gives
\begin{eqnarray}
\frac{dN_{q \bar q \to |q\bar q\rangle _{\rm dcnf}^{\rm res} \to l^+l^-}}{dM^2 d^4x} = N_c N_s^2 \sum_{f=1}^{N_f} \frac{\sigma_f^{\rm res}(M)}{2(2\pi)^4}  
\frac{\Gamma_{l^+l^-}}{\Gamma}
M^2\left ( 1-\frac{4m_q^2}{M^2} \right )^{\frac{1}{2}} T \sqrt{ \frac{\pi M T}{2} }e^{-M/T},  
\label{dilep1}
\end{eqnarray}
where the cross section $\sigma_f^{\rm res}(M)$ takes on the Breit-Wigner form 
\begin{eqnarray}
\sigma_f^{\rm res} (M)&=&\frac{4\pi}{3} \frac{Q_f^2(\alpha^\qedu)^2}{M^2} 
\frac{(\gamma/2)^2}{(E-E_{\rm res})^2)^2 +(\Gamma/2)^2}
\nonumber\\
 & & \times \left ( 1-\frac{4m_q^2}{M^2 }  \right )^ {-\frac{1}{2} }
 \left ( 1-\frac{4m_l^2}{M^2 }  \right )^ {\frac{1}{2} } \left( 1+2\frac{m_q^2+m_l^2}{M^2} +4 \frac{m_q^2 m_l^2}{M^4 }\right ),
 \label{sigma1}
\end{eqnarray}
The produced resonances have narrow widths $\Gamma$  which  can be presumed to fall within the experimental  bin with a  width of 2 MeV.  We obtain the theoretical  yield ratio 
  \begin{eqnarray}
\frac{{dN_{l^+l^- \to  |u\bar u \rangle_{\rm dcnf}^{\rm res} }}/dM }{dN_{l^+l^- \to |d\bar d \rangle_{\rm dcnf}^{\rm res} }/dM }
=\frac{ Q_u^2   \left ( 1-\frac{4m_l^2}{M^2 }  \right )^ {\frac{1}{2} } \left( 1+2\frac{m_q^2+m_l^2}{M^2} +4 \frac{m_q^2 m_l^2}{M^4 }\right ) M^{3/2}e^{-M/T}\}   |_{M=2m_u}                    }
            { Q_d^2   \left ( 1-\frac{4m_l^2}{M^2 }  \right )^ {\frac{1}{2} } \left( 1+2\frac{m_q^2+m_l^2}{M^2} +4 \frac{m_q^2 m_l^2}{M^4 }\right ) M^{3/2}e^{-M/T}\}     |_{M=2m_d}                    }.
\end{eqnarray}
This ratio of resonance yield   is dominated by the  quark charge ratio factor, $(Q_u/Q_d)^2=4$.  
By taking $m_u=2.5 \pm 0.90$ MeV and $m_d=4.95 \pm 1.85$ MeV  from the particle data group in \cite{PDG18},  we get the theoretical ratio to be 
\begin{eqnarray}
\frac{{dN_{l^+l^- \to  |u\bar u \rangle_{\rm dcnf}^{\rm res} }}/dM }{dN_{l^+l^- \to |d\bar d \rangle_{\rm dcnf}^{\rm res} }/dM }\bigg |_{\rm theory}
=4.03
\parbox{1cm}{+1.46\\ {}-1.07}.
\label{eq1}
\end{eqnarray}
 The total experimental $e^+e^-$ yield comes from both  on-resonance contributions and those from the
 off-resonance broad enhancement  contributions coming from the broad enhancement at 12 MeV with a FWHM of about 10 MeV.  We have earlier obtained a fit to this broad enhancement.  We can subtract out these  broad enhancement background  contributions to the yield contributions at the resonances at  $3\pm1$ MeV and $7\pm 1$ MeV and obtained the ratio of resonance yields,  
 \begin{eqnarray}
 \frac{{dN_{l^+l^- \to  |u\bar u \rangle_{\rm dcnf}^{\rm res}  }}/dM }{dN_{l^+l^- \to |d\bar d \rangle_{\rm dcnf}^{\rm res}  }/dM } \biggr |_{\rm experiment}
 &=& \frac{ {[{\rm Total - (Broad Bump Background)]}_{u \bar u}} }
 { {[{\rm Total - (Broad bump  Background)]}_{d \bar d}} } 
\nonumber\\
&=& \frac{ (2.85\pm 0.15) - 0} {(9.45\pm 0.6) - (7.47\pm 0.4) } = 1.44\parbox{1cm}{+0.81\\ -0.39},
\label{eq2}
\end{eqnarray}
which is close to the  theoretical predictions of Eq.\ (\ref{eq1}).  However, the above ratio depends sensitively on the shape  of  the broad enhancement near the threshold,  and there are threshold enhancement effects \cite{Won96a,Won96b,Sak48,Fad95,Sti95}, which will need to be considered 
 in future experimental analysis.

Within the uncertainties of the quark masses, we can therefore conclude that 
there are a good consistencies 
between theory and experiment 
 in resonance energies but the ratio of the production cross sections needs further analysis,  when we assign the  observed sharp resonances at $3\pm1$ MeV and $7\pm 1$ MeV to be the deconfined Coulomb-bound or color-Coulomb bound states of the deconfined quark and antiquark 
in 
 QED(U(1)-deconfined  color-singlet quark matter, with masses close to the quark rest masses.   The experimental evidence here  suggests  the tentative concept  that there may be new types of quark bound states or resonance in the deconfined color-singlet quark matter.

\subsection{Quark Confinement  in U(1) embedded in u
SU(3)$\times$U(1)/${\mathbb Z}_3$ 
in $(3+1)$D at $T=0$}
 
 In the temperature region $0\le T\le T_c(QED)$, neutral color-singlet quark matter should be confined and exist as a QED meson gas.  
 Our previous studies of QED mesons as confined in QED(U(1)) interaction in (3+1)D at $T=0$ utilized   a unified  product group  U(3)=SU(3)$\times$U(1)  with independent QCD(SU(3)) and QED(U(1)) interactions.  The product group $U(3)$   provides a valuable tool to examine
 QED mesons and  the dynamics of the quarks and gauge fields  \cite{Won10,Won20,Won23}
 with the success of the phenomenological open string model  of QCD and QCD mesons  \cite{Won10,Won20} and the the stretch (2+1)D flux tube model of quark confinement  in QED(U(1)) interaction in (3+1)D at $T=0$.  
 However,  these models needs  a well-founded theoretical justification for  quark confinement in (3+1)D at $T=0$.   Such a  simple product group U(3)=SU(3)$\times$U(1),  with an independent $U(1)$ would lead to the de-confinement of quarks in (3+1)D, as the strength of the coupling constant places quarks in the U(1) de-confinement regime, just as in electrons.   Furthermore, the  simple product group U(3)   also over-counts the combination of states and permits integer-charged quark  states which do not correspond to physical quark states in nature.
    
  To rectify these situations and to make progress towards   a  theoretical understanding of the possible quark confinement   in 
  QED(U(1))  in (3+1)D at $T=0$,
  we would like to examine quarks  in  the grand unification theory of quarks interacting in both QCD and QED interactions.  We follow  similar grand unification of the strong and electroweak interaction  pioneered by Georgi and Glashow \cite{Geo74}, Pati and Salam \cite{Pat74}.
 It is interesting to note however that for quarks,  whether the QCD(SU(3)) and the QED(U(1)) interactions can be unified is not yet fully explored.
 If  the QCD(SU(3)) interactoion and the QED(U(1) interaction can be unified   for quarks, they would share many properties in common, including perhaps the properties of quark confinement, deconfinement, 
and  phase transitions.  While there remains many experimental uncertainties in the grand unification theories
  as for example in the absence of unequivocal experimental evidence a  magnetic monopole \cite{Pre84}, the  study on  the implications of the grand unification of QCD and QED  on quark confinement may nonetheless be  one of the ways to find out theoretically the similarities and differences of the QCD and QED interactions and whether there may be theoretical and experimental supports for the grand unification for the QCD and the QED interactions.
 
 We shall show in Appendix A that in the grand unification theory of unifying the QCD and the QED interaction, quarks may be confined in the U(1) interaction that is embedded in the quotient group of 
 U(3)/${\mathbb Z}_3$ group, supporting the earlier phenomenological theory of the QED mesons in \cite{Won20} which we shall use.  A success of the phenomenological model will lent support for the unification of the QED(U(3)) and QED(U(1))  interactions for quarks.

To get a good determination of the QCD and QED meson masses, it is
necessary to take into account the flavor mixtures $D_{ij}^\lambda $
and the quark color and electric charges $Q_{\{u,d,s\}}^{\lambda}$,
the quark masses $m_f$, and the chiral condensate, as discussed in detail in 
\cite{Won20}.  We then obtain   the semi-empirical mass formula for a QED mesons with a flavor wave function
$\Phi_i$ given by
\begin{eqnarray}
\Phi_i=\sum_{f=1}^{N_f} D_{if} \phi_f,~~{\rm with~~} \phi_1=|u\bar u\rangle_{\rm cnf}^\qedu,  \phi_2=|d\bar d\rangle_{\rm cnf}^\qedu 
\end{eqnarray}
 can be estimated to be \cite{Won20}
\begin{eqnarray}
[m(\Phi_i)]^2=\frac{4\alpha^\qedu}{ \pi R_T^2} (\sum_{f=1}^{N_f} D_{if}Q_f^\qedu)^2  +
m_\pi^2\frac{\alpha^\qedu}{\alpha^\qcdu} \sum_{f=1}^{N_f} \frac{m_f}{m_{ud}} (D_{if})^2,
\label{qed}
\end{eqnarray}
where the first term on the right-hand side arises from the Schwinger
QED confinement mechanism generalized to two flavors, with
$Q_u^\qedu$=2/3, $Q_d^\qedu = -1/3$, $\alpha^\qedu=1/137$, and
$R_T=0.4 \pm 0.04$ fm \cite{Won20}. The second term on the right hand
side is the quark condensate contribution assumed to be proportional
to the coupling constant and calibrated to the pion mass $m_\pi$ for
which the first term from the Schwinger QED confinement mechanism
gives zero contribution for the QCD isovector state, and 
$\alpha^\qcdu=0.68\pm 0.08$ from earlier studies in \cite{Won20}.

The energy of the QED meson for the physical state $\Phi_i$ depends on
the mode of production and the isospin environmental constraint in
which the QED meson is produced.  It is not known what kinds of
isospin selection rule there may be when QCD(SU(3))-deconfined
color-octet and QED(U(1))-deconfined neutral color-singlet quark
matter need to obey, when they make the phase transition and freeze out.  For our
purposes, we shall consider all possibilities adhering to the strategy
that nature will avail itself of all possibilities that is not
explicitly prohibited.  Accordingly, we examine the QED mesons with
two flavors and isospin symmetry with good $I$ quantum numbers and
$I_z=0$, and we also examine cases in which the $q\bar q$ state
involves a quark with a single flavor and $I_z=0$.

QED mesons may also be produced in initial hadron-hadron collisions
prior to the formation of the quark gluon plasma, as revealed by the
production of anomalous soft photons in hadron-hadron collisions and
high-energy $e^+e^-$ annihilation.  They may also be produced during
the deconfinement to confinement phase transition of the quark
gluon plasma in an environment maintaining isospin symmetry involving
two flavors.  For QED((U(1))-confined systems with $I=\{0,1\}$ and
$I_z=0$, quantum numbers, the above mass formula (\ref{qed}) gives
isoscalar QED meson mass $m_{I=0,I_z=0}^\qedu=17.9\pm 1.5 $ MeV, and
isovector QED meson mass $m_{I=1,I_z=0}^\qedu=36.4 \pm 3.8 $ MeV
\cite{Won20}, as listed in Table 1.

In the confinement-to-deconfinement phase transition of the quark
photon plasma, the transition may occur in such an environment that
QED mesons may be formed with only a single flavor.  In such cases
with a single flavor the mass formula Eq.\ ({\ref{qed}}) gives theoretical QED
meson mass $m({|d\bar d \rangle}_{\rm cnf}^\qedu)=22.57\pm 2.2 $ MeV,
and QED meson mass $m({|u\bar u\rangle}_{\rm cnf}^\qedu)=33.69\pm 3.4
$ MeV, as listed in Table 1.
The experimentally resonance observed by \cite{Jai07} at 19$\pm$1,
may be interpreted as arising from isoscalar QED mesons with $I=0, I_z=0$.

 For completeness and for provisional  comparison 
between theory and experiment, we indicate  those weak resonances in Ref.\ \cite{Jai07} by a bracket on their resonance energies. 
It is encouraging that the energies of the weak experimental resonances at other energies in \cite{Jai07}
match those of the theoretical QED mesons of various type.  However, because of the low number of counts in the Pb-emulsion experiment, such matching could only be considered provisional.
It is hoped that future experiments with a great number of signal counts will allow a more significant comparison.

\subsection{QED Molecular States}

\begin{figure}[h]
\includegraphics[scale=0.80]{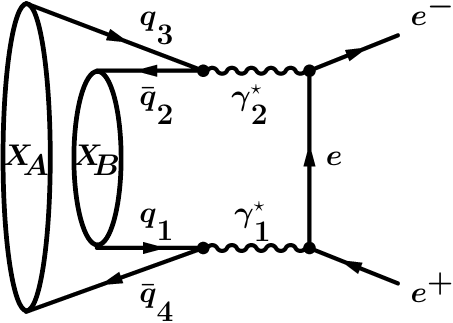} 
\vspace{0.8cm}
\caption{ Feynman diagram for the decay of the molecular state of [$|q
  \bar q\rangle_{\rm dcnf}^\qedu (X_A)\oplus|q \bar q\rangle_{\rm
    cnf}^\qedu (X_B)$], involving $|q\bar q\rangle$  of the same flavor,  into an $e^+e^-$ pair by the annihilation of the
  $q_3$ quark of $X_A$ with the antiquark $\bar q_2$ of $X_B$ and the
  annihilation of the antiquark $\bar q_4$ of $ X_A$ with the quark
  $q_1$ of $X_B$ into two virtual photons, which subsequently turn into
  an $e^+e^-$ pair. }
\label{mol}   
\end{figure}

We note that the
QED(U(1))-confined $|q \bar q\rangle_{\rm cnf}^\qedu$ state and the
QED(U(1))-deconfined $|q \bar q\rangle_{\rm dcnf}^\qedu$ state are
polar particles in the sense that they have electric charges of
opposite signs which interact with charge constituents of the other
particle through the long-range Coulomb interaction to form possible
molecular states as discussed in \cite{Won04}.  
The formation and detection of the molecular state will be facilitated 
when the 
constituents of the QED-deconfined bound state and the QED-confined QED meson state
involve quarks of the same flavor so that  their annihilation
into an $e^+e^-$ pair as indicated by the Feynman diagram in
Fig.\ \ref{mol} can occur.

In a molecular state, the molecular binding energy is small in
comparison with the masses of its molecular constituents (see for
example the molecular state in
Ref. \cite{Cho03,Won04,Tor04,Clo04,Swa04,Vol04}).  There is an approximate
semi-empirical mass equation for the occurrence of molecular states
\cite{Won22c}
\begin{eqnarray}
(\text{mass of a molecular state}) =\sum_i m_i - (\text{molecular binding energy}) \approx \sum_i m_i.
\end{eqnarray}

On the experimental side, there is a weak  observed $e^+e^-$ resonance at 29$\pm$1 MeV which does not seem to belong to any possible QED
mesons.  
It is interesting that  the experimental weak resonance state at 29$\pm$1 MeV  satisfies
approximately the necessary condition that        the resonance energy at 29$\pm$1 MeV is approximately
the sum of the mass of the $|d \bar d\rangle_{\rm dcnf}^\qedu$ state
and the mass of the $|d \bar d\rangle_{\rm cnf}^\qedu$ state,
\begin{eqnarray}
M(e^+e^-)|_{ {}_{29 \pm 1 {\rm ~MeV}}}\approx  M(|d \bar d\rangle_{\rm dcnf}^\qedu)|_{7\pm1 {\rm ~MeV}} 
+ M(|d \bar d\rangle_{\rm cnf}^\qedu)|_{23\pm1} {\rm ~MeV}.
\end{eqnarray}

We may tentatively  interpret the
experimental states at 29$\pm$1 as a possible molecular
state formed by $|q \bar q\rangle_{\rm cnf}^\qedu$ and $|q \bar
q\rangle_{\rm cdnf}^\qedu$ with $q=u,d$.  Statistically significant results however must await 
greater number of counts to allow a statistically significant comparison.

 If such an interpretation is
correct in future comparison, it involves the formation of the state during the phase
transition of both confined and deconfined phases, and it may be an
interesting probe on the dynamics during the phase transition.
Furthermore, the spatial sizes of the two states are quite different.
For example, the size of the QED meson state, $|d \bar d\rangle_{\rm
  cnf}^\qedu$, is of order $\hbar/(20 {\rm MeV}) \sim 10$ fm whereas
the size of the QED-deconfined bound state, $|d \bar d\rangle_{\rm
  dcnf}^\qedu$, is of order $\hbar/\alpha^\qedu m_d$, of order 100 fm.
The molecular state will likely to involve the greater size deconfined
$|d \bar d\rangle_{\rm dcnf}^\qedu$ state containing the smaller QED
meson $|d \bar d\rangle_{\rm cnf}^\qedu$ state in its interior.
Future investigation of such molecular states each involving a
QED-deconfined bound state and a QED-confined QED single-flavor meson
will be of great interest.

\section{The need for  Further Experimental  Investigations}

The comparison between the experiment Pb-emulsion data and theoretical
model results in Fig.\ 1 and Table I indicates that the highly complex
structure in the invariant mass spectrum of $e^+ e^-$ pairs from
high-energy Pb-emulsion collisions observed by Jain and Singh in
Ref. \cite{Jai07} may be explained as the production and the decay of
states of the neutral color-singlet $q\bar q$ quark matter in its deconfined
phase as well as in its confined phase.  However, the experimental
data are scanty, the nature of the spectrum with many resonances and
an enhancement is very complicated, the explanation offered to
interpret the experimental data is rather extraordinary involving
concepts that are novel in nature, and different measurements at
different laboratories gave slightly different resonance energies.
The theoretical model as well as the experimental background
subtraction also contain a high degree of uncertainties.  Data from new
experiments will provide new insight to assist future exploration
into the unknown.  There is therefore a need to carry out experimental
investigations such as those listed below to explore the uncharted new
territories:

\begin{enumerate}
\item 

 First and foremost should be the confirmation of the Pb-emulsion data
 on the complex structure of resonances and enhancements as observed
 by Jain and Singh \cite{Jai07} and El-Nagdy, Abdelsalam, and Badwady
 \cite{Eln07}.  It will be necessary to resolve the minor differences
 in resonance energies from different laboratories for many observed
 states.  It will also be necessary to improve on the statistics to
 confirm or refute the weaker resonances at higher masses.
  
 \item
 It will be useful to carry out similar high-energy heavy-ion-emulsion
 collision experiments with other projectiles at many different  energies to
 study the size and energy dependencies for the onset of QED(U(1)) and
 QCD(SU(3)) deconfinement.

 \item
As indicated in Table I and discussed in Section IIIE, the appearance
of new $q\bar q$ bound states of a quark and an antiquark in the new
deconfined mutual interactions provides a signature for the occurrence
of the deconfinement of neutral CSQM or COQM.  Furthermore, the bound state
energies will be modified from the purely Coulomb energy levels by the
screening of the charged medium.  Such a screening modification will
provide a handle to study the phenomenon of screening in quark photon
plasma or the quark gluon plasma \cite{Won96a,Won96b}.  It will be useful to study the
energies of these bound states experimentally.  In Table I, both the
theoretical QCD and the QED deconfined states occur at nearly the same
energies for the 3$\pm$1 MeV and the 7$\pm$1 MeV resonances.  The QCD
deconfined bound state and QED deconfined bound state have different
sizes and they have different lifetimes because their wave functions
at the origin are different.  They will decay at different distances
$L_0$ from the Pb-emulsion collision vertex.  By using the the
mean-free distances $L_0$ away from the Pb-emulsion vertex prior to
the $e^+e^-$ decay as an additional gate, one may try to resolve the
deconfined $|q\bar q\rangle_{\rm dcnf}^\qedu$ state from the $|q\bar
q\rangle_{\rm dcnf}^\qcdu$ deconfined bound states for the resonances
around 3$\pm1$ MeV and 7$\pm1$ MeV.  The measurements on the color-Coulomb bound
state energies of the deconfined COQM and the electric-Coulomb bound state
energies of the deconfined  neutral CSQM will provide useful information on
the onsets of deconfinement and the degree of color-charge screening
and electric-charge screening.

 \item
 In order to probe the QCD(SU(3))-deconfinement and the production of
 the quark gluon plasma, it will be useful to study experimentally the
 behavior of the $e^+e^-$ pair yield enhancement as a function of
 increasing $M(e^+e^-)$ in the region of $M(e^+e^-)$ beyond 50 MeV for
 the signal of $e^+e^-$ yield from the annihilation of quarks and
 antiquarks in a deconfined quark gluon plasma as described in
 Fig.\ \ref{ncoqm} in emulsion experiments.

\item
It will be beneficial to carry out high-energy AA collision
experiments to study the invariant mass distributions of
$\gamma\gamma$ pairs involving real photons or virtual photons in
order to search for the decays of the unknown $X$ particle states as
described in the Feynman diagrams in Figs. 2($d$), 2($e$) and 2($f$).
The lifetime of stable QED meson states X decaying into two real or
virtual photons in these diagrams varies as $1/(\alpha^\qedu)^6$ and
are $(1/\alpha^\qedu)^2$-times longer than the $e^+e^-$ decay lifetime
in diagram $2(c)$ which varies as $1/(\alpha^\qedu)^4$. If $e^+e^-$
pair decays in Diagram 2($c$) occur within $200\mu$m and gives the
results of Fig. 1, then such virtual or real photon decays as Diagrams
2($d$), 2($e$) and 2($f$) will occur at distances $L_0$ within $\sim
200$ $\mu$m$/(\alpha^\qedu)^2$ or 2.7m from the interaction vertex.
The counter experiments should be prepared to accept $e^+e^-$ pairs
which originate from emerging vertices at large distances from the AA
collision vertex.

\item 
Another important experiment is to confirm the observation of the
hypothetical E38 particle by carrying out experiments similar to those
performed at DUBNA \cite{Abr23}, by measuring the invariant masses of
$\gamma \gamma$ pairs with low invariant mass values  using high-energy
heavy ions on nuclear targets.  In particular, as the statistics for
the evidence for the hypothetical E38 is rather weak, there is a need
to improve on the number of counts in the detection of the E38
particle.

\item
The Feynman diagrams of Fig.\ 2($b$) and Fig.\ 2($c$) indicate that 
the 
decay of $q\bar q$ states of the confined or deconfined quark
matter into $e^+ e^-$ or into $\gamma \gamma$ have similar spectrum
such as those of Fig.\ 1 and Table I.  Already, the X17 resonance
detected in its $e^+e^-$ spectrum in high-energy Pb-emulsion
collisions of Fig.\ 1 is also found in the $\gamma\gamma$ spectrum in
the high-energy nucleus-nucleus collisions at DUBNA \cite{Abr23},
and the resonance at 7$\pm$1 MeV found in Fig.\ 1 can also be found  
in the $\gamma\gamma$ resonance at $\sim$9 MeV in Fig.\ 15 of \cite{Abr23}.  
 We
can predict that resonances with masses as those listed in Table I
should show up in the DUBNA-type experiments with the   invariant mass spectrum of $\gamma\gamma$ pairs.  It will be of great
interest to carry out such measurements of invariant mass spectrum of
$\gamma \gamma$ pairs at DUBNA or elsewhere.

\item
It will be useful to investigate in future anomalous soft photon experiments  \cite{Bai24}
whether  the production of the anomalous soft photons 
is indeed associated with the production of QED mesons by measuring the invariant masses of the $e^+e^-$ pairs.
We expect that  the $e^+e^-$ resonances such as those in Fig. 1  will show up as resonances in these measurements.  High-energy hadron-hadron, nucleus-nucleus, and $e^+$ - $e^-$ annihilation  collisions similar to those in earlier 
 anomalous soft production experiments in \cite{Chl84,Bot91,
  Ban93,Bel97,Bel02pi,Bel02,DEL06,DEL08,Per09,DEL10} should be carried out to search for both QED(U(1))-confined 
 and QED(U(1))-deconfined  $q\bar q$ states.

\end{enumerate}

\section{Summary, discussions, and conclusions}

In high-energy Pb-emulsion collision experiments carried out by Jain
and Singh \cite{Jai07}, the invariant mass spectrum of the $e^+e^-$
pairs emerging at vertices separated from the Pb-emulsion interaction
vertex at distances of 50 $\mu$m to 200 $\mu$m is highly complicated,
as shown in Fig.\ 1.  It
consists of resonances resting on top of an yield enhancement in the
anomalous mass region below 50 MeV.  
Some of the resonances have also been observed by other workers. 
The complexity
of the observed structure as well as the anomalous nature of their
masses made the search for its interpretation a difficult  task.
There has been no interpretation of these interesting results up to
the present time.

The classification scheme of the $q\bar q$ quark matter into different
color multiplet quark matters in confined and deconfined phases
proposed in \cite{Won26} provide the appropriate framework and
physical elements to accommodate reasonably well the main  features of the highly 
complex Pb-emulsion spectrum, including the low energy resonances and the enhancement of the
spectrum at $6 \lesssim M(e^+e^-$)$ \lesssim 20$ MeV, as shown in the
comparison in Fig.\ 1 and Table I.  In such a description, the
resonances at 3$\pm$1 and 7$\pm$1 MeV are described as arising from deconfined
down and up quarks and antiquarks interacting in the new interactions
in their QED(U(1))-deconfined  phase 
as an electric-Coulomb bound state or 
in their  QCD(SU(3))-deconfined phase as
a color-Coulomb bound state, with the resonance energies close to their
quark rest masses.  The resonance at 19$\pm$1 MeV is described as the
isoscalar QED(U(1))-confined QED meson and its energy places it to be a confirming
support for the existence of the hypothetical X17 particle observed by
ATOMKI \cite{Kra16}, DUBNA \cite{Abr23} and HUS \cite{Tra24}.  
The enhancement of the
spectrum at $6 \lesssim M(e^+e^-$)$ \lesssim 20$ MeV is described as arising from the annihilation of the
deconfined quarks and antiquarks  in the QED(U(1))-deconfined neutral color-singlet $q\bar q$ quark matter.

The approximate  agreement  
between the theory and the experimental spectrum suggests that both QED(U(1))-confined and QED(U(1))-deconfined
neutral color-singlet $q\bar q$ quark matter may have been produced in high-energy Pb-emulsion collisions, pending confirmation of the Pb-emulsion data.
 However, because of the scanty experimental
data,  the complicated  spectrum,  the extraordinary nature  of the explanation,
and the minor differences on resonance energies from different measurements, such a 
suggestion can only be  tentative in nature. 
It
necessitate the confirmation of the Pb-emulsion data by an independent laboratory.  More experimental measurements will also be needed to facilitate the exploration into unknown territories.

While we are waiting for the confirmation of the Pb-emulsion data, it
is of interest to contemplate on the implications of the main results
of the present  tentative findings.  If the Pb-emulsion data  turn out to be confirmed by further experimental works, then the present
exploration will open several new frontiers:
\begin{enumerate}
\item
A new form of matter. 

Neutral color-singlet $q\bar q$ quark matter would be a genuinely new phase,
distinct from the quark-gluon plasma that has been the focus of
heavy-ion physics for four decades. It interacts through
electromagnetic  rather than the strong force, placing it in a
qualitatively different regime.
\item
Independent support for X17. 

The X17 particle, first reported in 2016 in nuclear transitions at
ATOMKI, has become one of the most debated anomalies in particle
physics. The Pb-emulsion data provide evidence from a completely
different experiment, production mechanism, and energy scale. If both
experiments 
are seeing the same underlying QED meson, the case for its existence
is substantially strengthened.

\item
A window into QED confinement. 

The Schwinger mechanism for quark confinement via QED is well
established theoretically in (1+1) dimensions. Evidence that it also
operates in the real (3+1)-dimensional world for quarks and antiquarks
would be a major development in our understanding of gauge theory and
confinement.

\item

A connection to  the grand unification of QCD and QED interactions 

In Appendix A, we use the grand unification theory of QCD(SU(3)) and QED((U(1)) for quarks to justify  quark confinement in the U(1) interaction embedded in the grand unification theory of ${\cal G}=SU(3)\times U(1)/{\mathbb Z}_3$.  We already know that  the triplet nature of color is topologically connected with  the fractional charge of quarks.  
A confirmation of the Pb-emulsion data would 
be another indication of their common properties and possible roots as a unified interaction with a single coupling constant at a certain large mass scale.

\item
Connections to dark matter. 

In an earlier work \cite{Won20}, it was suggested that
self-gravitating assemblies of QED mesons, if sufficiently massive,
would not radiate $e^+$, $e^-$, or photons and would be good
candidates for part of the dark matter.  Similar considerations can
now be extended to the other resonances shown in Fig.\ 1, especially
the less massive ones such as the $| d\bar d\rangle_{\rm dcnf}$
resonance at 2-4 MeV, for which the requirements for being dark could
be satisfied with a less massive assembly.  Furthermore, the
color-singlet $dud$ QED neutron is stable under the QED(U(1)) interaction
while the corresponding $udu$ QED proton is unstable.  QED neutron therefore
cannot decay into a proton and can be another stable particle
candidate for the material of the dark matter \cite{Won22,Won22a}.

\end{enumerate}
Based on the above, the present paper promotes the exploration of physics on the uncharted frontier.

\vspace*{0.7cm}
\centerline{\bf Acknowledgments}
\vspace*{0.5cm} 
The author would like to thank Profs.\ Albert Wong,
Jack Yee Ng, A. Koshelkin,   T. Awes, S. Sorensen, I. Y. Lee, P. Zarubin, Y. Nakamura, 
T. Cs\"org\H o, H. Sadzjian,  D. Blaschke,
Che-Ming Ko,  Siu A. Chin,  Gang Wang, Huan Huang, Yunshan Cheng, and K. Abraamyan, for helpful discussions and
communications.  The research was supported in part by the Division of
Nuclear Physics, U.S. Department of Energy under Contract
DE-AC05-00OR22725.

\vspace*{0.7cm}

\appendix

 \section{ Quark Confinement in (3+1)D in QED(U(1)) at  $0\le T\le T_c(QED)$}
 
 We would like to emphasize on the outset  that the arguments presented in this Appendix are intended as  plausibility arguments rather than a first principle derivation for
 quark confinement in (3+1)D in QED(U(1)) at  $0\le T\le T_c(QED)$ . Even for the well-established case of quark confinement in QCD, no analytic proof from the continuum Yang-Mills Lagrangian exists and this is recognized as one of the Clay Mathematics Institute's Millennium Prize Problems \cite{Jaf00}. The mechanisms invoked here to explain confinement involving  dual superconductivity, monopole condensation, center vortices are themselves scenario-based pictures supported by lattice evidence rather than first-principles derivations, as unequivocal evidence for the magnetic monopole has also not been experimentally presented.   Lattice studies capable of testing such mechanisms directly, moreover, do not yet exist even for ordinary QCD in the chiral (near-massless quark) limit \cite{Due10,Tho02}, and no lattice formulation of the specific question of quark confinement in U(1)  embedded in SU(3)$\times$U(1) interactions  examined here has been attempted. Given this state of the field, we present the argument of this Appendix as a physically motivated plausibility case  for why the QED(U(1)) sector might confine quarks in the manner needed to explain Fig. 1 and Table I, rather than as a proof, and we regard a first-principle derivation  or lattice derivation as an important open problem for future work.

\subsubsection{Quark Confinement 
 in  a strong and electromagnetic grand unification theory in $(3+1)$D at $T=0$}

 The grand unification theory considers the unification of strong, electromagnetic, and weak interactions \cite{Geo74,Pat74}.  For our problem in the mass scales of 1-50 MeV, it suffices to limit ourselves only to the unification of the strong and the electromagnetic interactions and follow treatments pioneered by
 Polyakov \cite{Pol74,Pol87},  t'Hooft \cite{tHo74,tHo75}, Nielsen \cite{Nie73}, Nambu \cite{Nam74}, Mandelstam \cite{Man76}, and Greensite \cite{Gre03,Gre20}.
 The grand unification theory of quarks interacting in the QCD and QED interactions is described by  
the quotient group 
\begin{eqnarray}
{\cal G}(3)=\frac{U(3)}{{\mathbb Z}_3}=\frac{SU(3)\times U(1)}{{\mathbb Z}_3},
\end{eqnarray}
where   the center group  ${\mathbb Z}_3$
 of the product group $U(3)=SU(3)\times U(1) $,   
has elements $e^{i2\pi k/3} {\mathbb I}_{3\times 3}$ and $k=\{0, 1, 2\}$ in  the $SU(3)$ group together with a matching $e^{-2\pi k/3}$ phase shift element in the U(1) group,
\begin{eqnarray}
\{ e^{i2\pi k/3} {\mathbb I}_{3\times 3}, e^{-i2\pi k/3}\}\in {\mathbb Z}_3.
\end{eqnarray}
The above pairing
 ensures that the application of the center  element of ${\mathbb Z}_3$ on quarks leaves the quark state vector invariant.
  By the division of the group center ${\mathbb Z}_3$  containing the $SU(3)$ center elements with the corresponding matching  $U(1)$ element, the quotient group  topologically step-locks the 
color-charge degrees of freedom of quarks in the color  SU(3) group with the electric fractional charge degree of freedom of the quark in the electromagnetic U(1) group.   
By such a step-locking, the global glue ensures that any interaction field carrying a color representation in SU(3) interaction must possess matching fractional electric charges of (2/3) and (1/3) and vice versa, and is the topological origin of fractional charges \cite{Gel68}.    

It has been suggested that quantum chromodynamics and quantum electrodynamics may be unified at extremely short distances into a single gauge interaction 
 with a single coupling constant \cite{Geo74,Pat74}. Whether such a unification may be realized  at the grand unification scale remains to be investigated \cite{Pre84}. What is definitive is that the symmetry becomes spontaneously broken 
  in the region of our present interest, at around 1-50 MeV,
 leading to effective  QCD and the QED interactions with different coupling constants.
 If we treat the quarks interact in QED(U(1)) as an independent interaction with the coupling constant in the 1-50 MeV mass scale, the quarks will not be confined in the QED(U(1)) interaction.  
In contrast, if we now study 
quark confinement in (3+1)D in the QED(U(1)) interaction at T=0 embedded  in the  strong and electromagnetic grand unification theory and if such theory leads to quark confinement   in (3+1)D in U(1) as QED mesons at T=0, 
then theoretical and experimental evidence of the Pb-emulsion data suggest a strong case for the unification of SU(3) and U(1) interaction for quarks.

The grand unification of strong and electromagnetic interaction  possesses  many interesting  properties.   Embedding U(1) inside 
U(3)/${\mathbb Z}_3$ restricts the Abelian component to the circle group of $S^1$   which yields  the first homotopy group $\pi_1(S^1)={\mathbb Z}$, unlocking the magnetic monopole instantons which power the Polyakov sector of QED U(1) confinement in (1+1)D and (2+1)D \cite{Pol87}.  The ${\mathbb Z}_3$ of both SU(3) and U(1) explicitly step-lock the non-Abelian color transformation to the Abelian phase shift transformation and ensures that any color  field carrying a color representation must possess the matching fractional charge.  It provides the topological origin of the fractional electric charges \cite{Gel68}.  The grand unification  also provides the regularization of the electromagnetic interaction at short distances, as the instanton action of the magnetic monopoles is bounded by the SU(3) unification scale.   
 \cite{Geo74,Pat74}
\subsubsection{ Monopoles in Grand Unification Theories}

As shown by t'Hooft\cite{tHo74} and Polyakov \cite {Pol74,Pol87}, all grand unification theories possess magnetic monopoles. 
  We can therefore study the behavior of the magnetic monopole 
to understand why quarks may be  confined in $U(1)$ interactions embedded inside the quotient group $\cal G$=$U(3)/{\mathbb Z}_3$ framework, while electrons in U(1) interaction remain deconfined  in (3+1)D at $T=0$. 
We shall follow Polyakov \cite{Pol74,Pol87}, t'Hooft \cite{tHo74,tHo75}
Nielsen \cite{Nie73},   Nambu \cite{Nam74},   Mandelstam \cite{Man76}, and Greensite\cite{Gre03,Gre20},  and shall examine  whether  the condensation of magnetic monopoles  in the transverse direction  leads  to formation of   a  flux tube on which the Schwinger mechanism may become operative for a valence color-singlet $q\bar q$
 pair in U(1) as a  QED mesons in (3+1)D at $T=0$.

In a conventional random $(3+1)\text{D}$ geometry, magnetic monopoles are point-like in space but form closed loop trajectories of dimension $d=1$ in Euclidean spacetime, governed by the mapping $\pi_2(S^2) = \mathbb{Z}$ on a 3-dimensional hypersurface. The action of a loop of perimeter $L$ scales as $S_{\text{loop}} \sim M_M L$, where $M_M \sim 4\pi / g^2$ is the monopole mass. The configuration entropy scales as $S_{\text{entropy}} \sim \ln(\mu) L$. For a weak coupling $g^2 < 1$, $S_{\text{loop}} > S_{\text{entropy}}$, forcing the loop fugacity to vanish identically in the continuum limit and thus electrons are deconfined.    Hence, electric charge are de-confined in (3+1)D if they interact only in U(1) interaction.

\subsubsection{Construction of the Compact Partition Function}

We begin by defining the Euclidean partition function for the ${SU(3) \times U(1)}/{\mathbb{Z}_3}$ quotient group. The division by the center subgroup $\mathbb{Z}_3$ enforces a global topological constraint that locks the $SU(3)$ color states with fractional $U(1)$ electric charges. This allows for compact $U(1)$ configurations that are fundamentally linked to the $SU(3)$ root. 

The total Euclidean action $S$ is expressed as:
\begin{equation}
S = \int d^4x \left[ \frac{1}{4g_{SU(3)}^2} (F_{\mu\nu}^a)^2 + \frac{1}{4g_{U(1)}^2} (f_{\mu\nu})^2 \right] + S_{\text{top}},
\end{equation}
where $F_{\mu\nu}^a$ is the $SU(3)$ field strength tensor, $f_{\mu\nu}$ is the $U(1)$ field strength, and $S_{\text{top}}$ represents the topological terms governing instanton boundary configurations. 

We consider now a  valence quark-antiquark ($q\bar{q}$) pair with one flavor as an example separating along the $z$-axis ($x^3$) in a $(3+1)\text{D}$ Minkowski spacetime. The strong-coupling infrared dynamics of the underlying background $SU(3)$ sub-sector squeeze the total chromatic energy into a cylindrical flux tube of radius $R_{\perp} \sim \Lambda_{\text{QCD}}^{-1}$. 

These background fields restrict the kinematics of the embedded $U(1)$ gauge field $A_{\mu}$. For distances $|z| \gg R_{\perp}$, the gauge fields can be decomposed into longitudinal components $A_a$ (where $a \in \{0, 3\}$) and transverse components $A_i$ (where $i \in \{1, 2\}$). The effective action factors into a longitudinal world-sheet term and a transverse boundary term:
\begin{equation}
S = \int d^4x \left[ -\frac{1}{4} F_{\mu\nu}F^{\mu\nu} \right] \quad \xrightarrow{\text{Infrared Limit}} \quad S_{\text{long}} + S_{\text{trans}}.
\end{equation}

We wish to show that because of the presence of interaction of the underlying monopoles,  the longitudinal electric flux is localized to the tube core and the field strength components $F_{01}$ and $F_{02}$ vanish exponentially at large transverse distances. The longitudinal sector then reduces to an effective two-dimensional Schwinger model on the world-sheet $\mathcal{M}_2 = \mathbb{R}^{1,1}$:
\begin{equation}
S_{\text{long}} = \pi R_{\perp}^2 \int_{\mathcal{M}_2} d^2x \, \left[ -\frac{1}{4} F_{ab}F^{ab} + e_{\text{eff}} A_a j^a \right],
\end{equation}
where $g_{\text{eff}} =g_\2d= g_\4d / (\sqrt{\pi} R_{\perp})$ is the geometrically enhanced effective coupling constant \cite{Won09}.

\subsubsection{ Transverse Polyakov Projection and Monopole Point Defects in (2+1)D}

Inside the $U(3)$ flux tube, the $(1+1)\text{D}$ longitudinal pinning modifies the topological boundary conditions. The remaining spatial fluctuations of the compact $U(1)$ gauge field are projected onto the transverse spatial slices synchronized with Euclidean time: $\mathcal{M}_3 = \mathbb{R}^2 \times S^1_{\beta}$. 

On this restricted $(2+1)\text{D}$ manifold, the topological defects are no longer dynamic loops, but rather point-like instantons (Polyakov monopoles) centered at positions $x_i = (\mathbf{x}_{\perp}, \tau)$. The homotopy mapping becomes:
\begin{equation}
\pi_1(U(1)_{\text{compact}}) = \pi_1(S^1) = \mathbb
{Z}.
\end{equation}
The magnetic field configuration for a localized defect at the origin of $\mathcal{M}_3$ is given by the dual identity:
\begin{equation}
B_{\mu} = \frac{1}{2}\epsilon_{\mu\nu\lambda}F^{\nu\lambda} = \frac{m}{2 e} \frac{x_{\mu}}{|x|^3}, \quad \mu,\nu,\lambda \in \{1,2,4\},
\end{equation}
where $m \in \mathbb{Z}$ is the magnetic charge winding number. Because these defects are point-like in $(2+1)\text{D}$, their classical action configuration is strictly localized and independent of any loop perimeter:
\begin{equation}
S_{\text{point}} = \frac{4\pi^2}{g_{SU(3)}^2}  \kappa,
\end{equation}
where $\kappa$ is a dimensionless geometric regularizer determined by the flux tube radius $R_{\perp}$.

\subsubsection{ Non-Zero Monopole Fugacity and Plasma Condensation}
Because the topological defects are point-like monopole  instantons rather than macroscopically extended loops, their grand canonical partition function corresponds to a classic classical Coulomb gas of magnetic monopoles. The effective monopole fugacity $\zeta_{SU(3)}$ is strictly non-zero:
\begin{equation}
\zeta_{SU(3)} = \exp\left( -S_{\text{point}} \right) = \exp\left( -\frac{4\pi^2 \kappa}{g_{SU(3)}^2} \right) > 0.
\end{equation}
Even in the weak-coupling regime ($g^2 \to 0$), $\zeta$ remains structurally bounded away from zero because the non-Abelian background screening restricts $\kappa \to 0$ as the gauge group unifies at the $U(3)$ scale.

The total magnetic current density operator is expressed as a sum over point sources: $j_M(x) = \sum_a m_a \delta^{(3)}(x - x_a)$. Performing the standard duality transformation maps the compact $U(1)$ sector to a scalar field $\chi(x)$ representing the magnetic potential. The sine-Gordon effective Lagrangian density on the $\mathcal{M}_3$ slice reads:
\begin{equation}
\mathcal{L}_{\text{eff}} = \frac{1}{2}(\partial_{\mu}\chi)^2 - 2\zeta_{SU(3)} \cos\left(\frac{4\pi}{g_{SU(3)}}\chi\right).
\end{equation}
Expanding the cosine potential around its global minimum yields the dynamically generated mass gap for the dual photon:
\begin{equation}
m_{\text{dual}}^2 = \zeta_{SU(3)} \left(\frac{32\pi^2}{g_{SU(3)}^2}\right) > 0.
\end{equation}
Because $m_{\text{dual}} > 0$, the magnetic monopole gas behaves as a screening plasma. By the dual Meissner effect, this plasma expels any remaining transverse electric flux components, compressing the field lines strictly into the $(1+1)\text{D}$ core.

This hybrid geometric mechanism demonstrates that the $U(1)$ sector of valence quarks cannot undergo standard $(3+1)\text{D}$ deconfinement. The non-Abelian $SU(3)$ core provides the physical $(1+1)\text{D}$ Schwinger flux pipe, while the compact boundary conditions map the Abelian degrees of freedom to a $(2+1)\text{D}$ Polyakov instanton gas. This ensures that the monopole fugacity $\zeta_{SU(3)}$ remains non-vanishing, enforcing a linear confinement potential $V(r) = \sigma r$ inside the quark sector that is geometrically blocked for standard leptons.

\subsubsection{Topological Locking in the $SU(3) \times U(1)/\mathbb{Z}_3$ quotient group}

This global topological gluing ensures that the magnetic sectors of the respective fields cannot vary independently. We define  $\vec{\phi} = (\phi_1, \phi_2)$ to be  the dual photon fields resolving the abelianized magnetic sectors of $SU(3)$, and let $\sigma$ define the dual field parameterizing the compact $U(1)$ sector. The low-energy effective dual Lagrangian density $\mathcal{L}_{\text{dual}}$ in $(2+1)\text{D}$ Euclidean space-time for these transverse fields is governed by:
\begin{equation}
\mathcal{L}_{\text{dual}} = \frac{1}{2} (\partial_i \vec{\phi})^2 + \frac{1}{2}(\partial_i \sigma)^2 - \zeta_{SU(3)} \sum_{k=1}^{3} \cos\left( \vec{\alpha}_k \cdot \vec{\phi} + \frac{2}{3}\sigma \right),
\end{equation}
where $\vec{\alpha}_k$ represent the root vectors of $SU(3)$, and $\zeta_{SU(3)}$ represents the instanton gas fugacity dictated by non-perturbative topological determinantal pre-factors. Due to the physical constraint where the strong sector coupling dominates the electromagnetic sector ($\alpha_{SU(3)} \gg \alpha_{U(1)}$), the fugacity and overall energy scales are dictated strictly by the non-perturbative $SU(3)$ vacuum. The weak $U(1)$ electromagnetic field $\sigma$ is locked directly into the potential minimum due to the exact $\mathbb{Z}_3$ topological identification, forcing it to undergo identical vacuum configurations as the color fields.

\subsubsection{Geometrical Invariance of the Polyakov Flux Tube}
The dual Meissner effect drives the confinement mechanism in this $(2+1)\text{D}$ projection. Because the monopole gas condenses, it expels the longitudinal gauge electric fields, compressing them into a localized string or flux tube along the longitudinal direction. 

Expanding the potential around its vacuum expectation value yields a uniform screening mass matrix. Using the standard normalization $\sum_{k=1}^3 \alpha_{k,i}\alpha_{k,j} = \frac{3}{2}\delta_{ij}$, the dual photon fields acquire a mass gap given by:
\begin{equation}
m_{\text{dual}} = \sqrt{\frac{3}{2}\zeta_{SU(3)}}.
\end{equation}
The transverse radius $R_T$ of the resulting flux tube is governed by the inverse of this dual photon mass, yielding:
\begin{equation}
R_T \sim \frac{1}{m_{\text{dual}}} = \sqrt{\frac{2}{3\zeta_{SU(3)}}}.
\end{equation}
Because $R_T$ is entirely determined by the underlying vacuum instanton density $\zeta_{SU(3)}$, the physical profile of the flux tube is a structural property of the compressed vacuum background rather than a property of the valence particles running through it. Consequently, when a valence quark-antiquark pair ($q\bar{q}$) is produced in a color-singlet or a color-octet  configuration, the geometric boundaries of the channel remain constant at $R_T \approx 0.4\text{ fm}$ regardless of whether the active longitudinal interaction is mediated by the QCD $SU(3)$ force or the QED $U(1)$ force \cite{Won20}. The physical channel does not dilate or contract because its structural integrity is enforced by the topological locking of the background monopole gas.

With the transverse degree of freedom localized within a fixed radius $R_T$ as determined by the underlying (and hidden) color vacuum that is dominated by the SU(3) interaction, the longitudinal field dynamics of the valence quark and antiquark along the string axis can be mapped to an effective $(1+1)\text{D}$ space-time strip. Integrating the field energy over the fixed cross-sectional area of the flux tube $A = \pi R_T^2$ yields an effective coupling constant  that scales linearly with (3+1)D  gauge coupling as carried out  in 
\cite{Won09,Won23}, we obtain 
\begin{eqnarray}
(g_{\2d})^2=\frac{1}{\pi
    R_T^2}(g_{\4d})^2=\frac{4\alpha_{\4d}}{R_T^2}.
\label{12}
\end{eqnarray}
which allows a transcription of the open-string description in (1+1)D to a
flux tube description in (3+1)D for the valence  quark and antiquark interacting in QCD as a QCD interactions.
The dynamics of the flux tube is then reduced into that of the Schwinger model of (1+1)D which exhibits confinement.   
 By applying the massless Schwinger mechanism, the generated pseudo-scalar boson mass 
 for the valence quark and antiquark, the mass of the  the QED and QCD mesons are then 
 \begin{eqnarray}
m_\qcd^2=\frac{(g_\2d^\qcd)^2}{\pi}=\frac{4 \alpha_\qcd}{\pi R_T^2},~~~~~~~ m_\qed^2=\frac{(g_\2d^\qed)^2}{\pi}=\frac{4 \alpha_\qed}{\pi R_T^2}.
\label{eq19}
\end{eqnarray}
With $\alpha_{\4d}^{\qed}$\!=$\alpha_{{}_{\rm QED}}$=1/137,
$\alpha_{\4d}^{\qcd}$\!=$\alpha_s $ $\sim$ 0.68 from hadron
spectroscopy \cite{Won00,Won01} and $R_T$$\sim$0.4 fm from lattice QCD
calculations \cite{Cos17} and $\langle p_T^2 \rangle $ of produced
hadrons in high-energy $e^+e^-$ annihilations \cite{Pet88}, we
estimate the masses of the open string QCD and QCD mesons to be
\begin{eqnarray}
m^{\qcd}\sim 458{\rm ~ MeV}, ~~~ {\rm and}~~ m^{\qed}\sim 48 {\rm
  ~MeV}.
\label{eq23}
\end{eqnarray}
This derivation using a single-flavor leads to quark confinement  provides a rigorous field-theoretic foundation for 
quark confinement  in the QED(U(1))  embedded in the grand unified interaction of ${\cal G}=SU(3)\times U(1)/{\mathbb Z}_3$ interaction.  
It also justifies 
the fixed-radius assumption utilized successfully in prior phenomenological works to account for both standard QCD meson spectra and QED meson states \cite{Won20}.  It can therefore be used to examine 
the resonances  observed in Pb-emulsion collisions.

\end{document}